\documentclass[aps,pra,twocolumn,showpacs,floatfix]{revtex4}
\usepackage{graphicx}
\usepackage{nicefrac}
\usepackage{amsmath}
\usepackage{amsfonts}
\usepackage{amssymb}
\usepackage{amsthm}
\usepackage{epsf}
\usepackage{bm}
\usepackage{bbm}
\usepackage{longtable}

\usepackage{dcolumn}

\newcolumntype{.}{D{x}{}{-1}}

\newcommand{\bsigma}{\vec{\sigma}}

\newcommand{\bfp}{\vec{p}}

\newcommand{\bfr}{\vec{r}}

\newcommand{\Za}{{Z\alpha}}

\newcommand{\cE}{{\cal E}}

\newcommand{\lbr}{\left<} \newcommand{\rbr}{\right>}

\begin{document}

\title{Theoretical energies of low-lying states of light helium-like ions}

\author{Vladimir A. Yerokhin} \affiliation{Center for Advanced Studies,
        St.~Petersburg State Polytechnical University, Polytekhnicheskaya 29,
        St.~Petersburg 195251, Russia}

\author{Krzysztof Pachucki} \affiliation{Institute of Theoretical
        Physics, University of Warsaw, Ho\.{z}a 69, 00--681 Warsaw, Poland}

\begin{abstract}
  Rigorous quantum electrodynamical calculation is presented for energy
  levels of the $1^1S$, $2^1S$, $2^3S$, $2^1P_1$, and $2^3P_{0,1,2}$ states of
  helium-like ions with the nuclear charge $Z=3\ldots 12$. The calculational
  approach accounts for all relativistic, quantum electrodynamical, and recoil
  effects up to orders $m\alpha^6$ and $m^2/M\alpha^5$, thus advancing the
  previously reported theory of light helium-like ions by one order
  in $\alpha$.
\end{abstract}


\maketitle

\section{Introduction}

Atomic helium and light helium-like ions have long been attractive subjects
of theoretical and experimental investigations. From the theoretical
point of view, helium-like atoms are the simplest few-body systems. As
such, they are traditionally used as a testing ground for different 
methods of the description of atomic structure. On the experimental side,
small natural linewidths of transitions between the metastable $^3P$ and
$^3S$ states of helium-like ions 
permit spectroscopic measurements of high precision. For atomic
helium, experimental investigations are nowadays carried out with the relative
accuracy up to $7\times 10^{-12}$ \cite{pastor:04}. An advantage of the
helium-like ions as compared to, e.g., hydrogen-like ones, is that the transition
frequency increases slowly with the nuclear charge number $Z$ ($\sim Z$). 
This feature ensures that wavelenghts of a significant part of the helium 
isoelectronic sequence fall in the region suitable for accurate experimental 
determination.

There are presently two main theoretical approaches that allow one to
systematically account for the electron-correlation, relativistic, and quantum
electrodynamical (QED) effects in few-electron systems.  The first one,
traditionally used for light systems, relies on an expansion of the
relativistic and QED effects in terms of $\alpha$ and $\Za$ ($\alpha$ is the
fine-structure constant) and treats the nonrelativistic electron-electron 
interaction nonperturbatively.  This approach started with 
the pioneering works of Araki \cite{araki:57} and Sucher
\cite{sucher:58}, who derived the expression for the Lamb shift in
many-electron systems complete through the order $m\,\alpha^5$. 
The other approach aims primarily at high-$Z$ ions. It does not use
any expansion in the binding-strength parameter $\Za$ (and thus is often
referred to as the {\em all-order} approach) but treats the electron-electron
interaction within the perturbative expansion with the parameter $1/Z$. A
systematic formulation of this method is presented in
Ref.~\cite{shabaev:02:rep}.

These two approaches can be considered as complementary, the first being
clearly preferable for light atoms and the second, for heavy ions. The
intermediate region of nuclear charges around $Z=12$ is the most difficult one
for theory, as contributions not (yet) accounted by
either of these methods have their maximal value there. In order to provide
accurate predictions for the whole isoelectronic sequence, it is
necessary to combine these two approaches.

For the first time a combination of the complementary approaches was made
by Drake \cite{drake:88:cjp}. His results for energies of
helium-like ions comprise all effects up to order $m\alpha^5$ 
in the low-$Z$ region,
whereas in the high-$Z$ region, they are complete up
to the next-to-the-leading order in $1/Z$ for nonradiative effects and to the
leading order, for radiative effects.
Since then, significant
progress was achieved in theoretical understanding of energy levels of atomic
helium, whose description is now complete through order
$m\alpha^6$ \cite{pachucki:06:hesinglet,pachucki:06:he}.  Also in the high-$Z$
region, theoretical energies have recently been significantly
improved by a rigorous treatment of the two-electron QED corrections
\cite{artemyev:05:pra}, which completed the $O(1/Z)$ part of the radiative effects.

In the present investigation we aim to improve theoretical predictions of
the $n=1$ and $n=2$ energy levels of light helium-like ions.  To this end, we
perform a calculation that includes all QED and recoil effects up to
orders $m\alpha^6$ and $m^2/M\,\alpha^5$ ($M$ is the nuclear mass).  
In order to establish a basis for
merging the current approach with the all-order calculations, we perform an
extensive analysis of the $1/Z$ expansion of individual corrections. This 
analysis also
provides an effective test of consistency of our calculational results and of the
$1/Z$-expansion data available in the literature.

\section{Theory of the energy levels}

In this section, we present a summary of contributions to the
energy levels of two-electron atoms complete up to orders $m\alpha^6$ and
$m^2/M\,\alpha^5$.

According to QED theory, energy levels of atoms are represented by an
expansion in powers of $\alpha$ of the form
\begin{equation} \label{e1} E(\alpha) = E^{(2)} + E^{(4)} + E^{(5)} + E^{(6)}
  + E^{(7)} + \ldots,
\end{equation}
where $E^{(n)}\equiv m\alpha^n\cE^{(n)}$ is a contribution of order $\alpha^n$
and may include powers of $\ln\alpha$. Each of $\cE^{(n)}$ is in turn expanded
in powers of the electron-to-nucleus mass ratio $m/M$
\begin{equation} \label{e2} \cE^{(n)} = \cE^{(n)}_{\infty}+ \cE^{(n)}_M +
  \cE^{(n)}_{M^2}+\ldots\,,
\end{equation}
where $\cE^{(n)}_M$ denotes the correction of first order in $m/M$ and
$\cE^{(n)}_{M^2}$ is the second-order correction.  To note, for the
nonrelativistic energy, it is more natural to expand in $m_r/M$ (where $m_r$
is the reduced mass) rather than in $m/M$, since such expansion has smaller
coefficients. For the relativistic corrections, however, the natural recoil
expansion parameter is $m/M$, so for consistency we use it for the
nonrelativistic energy as well.

The terms of the double perturbation expansion (\ref{e1}) and (\ref{e2}) 
are expressed as expectation values of some effective Hamiltonians (in some
cases, of nonlocal operators) and as second- and higher-order perturbation
corrections induced by these Hamiltonians (operators). It is noteworthy that
the expansion (\ref{e1}) is employed also for the states that are mixed by the
relativistic effects, namely $2^1P_1$ and $2^3P_1$.
The mixing effects are treated perturbatively. (So, the leading effect
due to the $2^1P_1-2^3P_1$ mixing appears naturally as the second-order
$m\alpha^6$ correction, together with contributions from other intermediate
states.) This differs from the approach used, e.g., in
Ref.~\cite{drake:88:cjp}, where a two-by-two matrix was constructed for this
pair of states and the energies were obtained by a diagonalization.

The leading contribution to the energy $\cE^{(2)}_{\infty}\equiv \cE_0$ is the
eigenvalue of the nonrelativistic Hamiltonian,
\begin{equation}
  H^{(2)} \equiv H_0 = \sum_a \biggl(\frac{\vec p_a^{\,2}}{2} -
  \frac{Z}{r_a}\biggr) + \sum_{a<b} \frac1{r_{ab}}.
\end{equation}
The first- and second-order recoil corrections to the nonrelativistic energy
are given by
\begin{align}
  \cE^{(2)}_M = -\frac{m}{M}\, \cE^{(2)}_{\infty} + \left< H^{(2)}_{{\rm rec}}
  \right>\,,
\end{align}
\begin{align}
  \cE^{(2)}_{M^2} = &\ \left(\frac{m}{M}\right)^2\, \cE^{(2)}_{\infty} -2 \,
  \frac{m}{M}\,\left< H^{(2)}_{{\rm rec}} \right> \nonumber \\ & +\left<
    H^{(2)}_{{\rm rec}} \frac1{(\cE_0-H_0)'} H^{(2)}_{{\rm rec}}\right>\,,
\end{align}
where
\begin{equation} \label{MP} H^{(2)}_{{\rm rec}} = \frac{m}{M}\,
  \sum_{a<b}\vec{p}_a\cdot\vec{p}_b
\end{equation}
is the mass polarization operator.

The leading relativistic correction $\cE^{(4)}_{\infty}$ is given by the
expectation value of the Breit-Pauli Hamiltonian $H^{(4)}$
\cite{bethesalpeter},
\begin{align} \label{H4} H^{(4)} &\ =\sum_a \biggl[-\frac{\vec p^{\,4}_a}{8} +
  \frac{ \pi Z}{2}\,\delta^3(r_a) +\frac{Z}{4}\, \vec\sigma_a\cdot\frac{\vec
    r_a}{r_a^3}\times \vec p_a\biggr]
  \nonumber \\
  & +\sum_{a<b}\biggl\{ -\pi\, \delta^3(r_{ab}) -\frac1{2}\, p_a^i\,
  \biggl(\frac{\delta^{ij}}{r_{ab}}+\frac{r^i_{ab}\,r^j_{ab}}{r^3_{ab}}
  \biggr)\, p_b^j \nonumber \\ & - \frac{2 \pi}{3}\,\vec\sigma_a
  \cdot\vec\sigma_b\,\delta^3(r_{ab}) +\frac{\sigma_a^i\,\sigma_b^j} {4
    r_{ab}^3}\,
  \biggl(\delta^{ij}-3\,\frac{r_{ab}^i\,r_{ab}^j}{r_{ab}^2}\biggr) \nonumber
  \\ & +\frac1{4\,r_{ab}^3} \bigl[ 2\,\bigl(\vec\sigma_a\cdot\vec
  r_{ab}\times\vec p_b - \vec\sigma_b\cdot\vec r_{ab}\times\vec p_a\bigr)
  \nonumber \\ & + \bigl(\vec\sigma_b\cdot\vec r_{ab}\times\vec p_b -
  \vec\sigma_a\cdot\vec r_{ab}\times\vec p_a\bigr)\bigr]\biggr\}\,.
\end{align}
  
The finite nuclear mass correction to the Breit contribution $\cE^{(4)}_M$ is
conveniently separated into the mass scaling, the mass polarization, and the
operator parts. The mass scaling prefactor is $(m_r/m)^4$ for the first term
in Eq.~(\ref{H4}) and $(m_r/m)^3$, for all the others. The mass polarization
part represents the first-order perturbation of $\cE^{(4)}_{\infty}$ by the
mass-polarization operator (\ref{MP}). The operator part is given by the
expectation value of the recoil addition to the Breit-Pauli Hamiltonian,
\begin{align} \label{fsrec} H^{(4)}_{\rm rec} = &\ \frac{Zm}{2M}\, \sum_{ab}
  \biggl[\frac{\bfr_a}{r_a^3}\times \bfp_b \cdot\bsigma_a -
  p_a^i\,\left(\frac{\delta^{ij}}{r_a}+\frac{r^i_ar^j_a}{r_a^3}\right) p_b^j
  \biggr] \,.
\end{align}

$\cE^{(5)}_{\infty}$ is the leading QED correction \cite{araki:57,sucher:58}.
We divide it into the logarithmic and the nonlogarithmic parts,
$\cE^{(5)}_{\infty} = \cE^{(5)}_{\infty}({\rm log})+\cE^{(5)}_{\infty}({\rm
  nlog})$, which are given by
\begin{align} \label{E5log} \cE^{(5)}_{\infty}({\rm log}) &\ = \frac{14}{3}\ln
  (\Za)\, \sum_{a<b} \langle \delta^3(r_{ab}) \rangle \nonumber \\ & +
  \frac{4\,Z}{3} \, \ln \left[(\Za)^{-2}\right]\, \sum_a \langle \delta^3(r_a)
  \rangle\,,
\end{align}
and
\begin{align} \label{E5nlog} \cE^{(5)}_{\infty}({\rm nlog}) &\ =
  \frac{164}{15} \sum_{a<b} \langle \delta^3(r_{ab}) \rangle
  -\frac{14}{3}\,\sum_{a<b}\widetilde{Q}_{ab} \nonumber \\ &
  +\left[\frac{19}{30}-\ln \left( \frac{k_0}{Z^2}\right) \right] \frac{4Z}{3}
  \sum_a\langle \delta^3(r_a) \rangle + \langle H_{\rm fs}^{(5)} \rangle\,,
\end{align}
where
\begin{align} \label{Q} \widetilde{Q}_{ab} = \left< \frac1{4\pi
      r_{ab}^3}+\delta^3(r_{ab})\,\ln Z \right>\,,
\end{align}
and the singular operator $r^{-3}$ is defined by
\begin{align} \label{rcube} \left\langle\frac{1}{r^3}\right\rangle &\ \equiv
  \lim_{a\rightarrow 0}\int d^3 r\, \phi^{*}(\vec r)\,\phi(\vec r) \nonumber
  \\ & \times \left[\frac{1}{r^3}\,\Theta(r-a) + 4\,\pi\,\delta^3(r)\,
    (\gamma+\ln a)\right]\,,
\end{align}
where $\gamma$ is the Euler constant.  The Bethe logarithm is defined as
\begin{align} \label{bethe} \ln (k_0) = \frac{\Bigl\langle\sum_a \vec
    p_a\,(H_0-\cE_0)\, \ln\bigl[2\,(H_0-\cE_0)\bigr]\, \sum_b\vec
    p_b\Bigr\rangle}{2\,\pi\,Z\, \Bigl\langle\sum_c\delta^3(r_c)\Bigr\rangle}.
\end{align}
The operator $H_{\rm fs}^{(5)}$ is the anomalous magnetic moment correction to
the spin-dependent part of the Breit-Pauli Hamiltonian. $H_{\rm fs}^{(5)}$
does not contribute to the energies of the singlet states and to the
spin-orbit averaged levels but it yields the $m\alpha^5$ contribution to the
fine structure splitting. It is given by
\begin{align} \label{H5fs} H^{(5)}_{\rm fs} &\ = \frac{Z}{4\pi}\,\sum_a
  \vec\sigma_a\cdot\frac{\vec r_a}{r_a^3}\times \vec p_a \nonumber \\ &
  +\sum_{a<b}\biggl\{ \frac1{4\,\pi}\frac{\sigma_a^i\,\sigma_b^j} {r_{ab}^3}\,
  \biggl(\delta^{ij}-3\,\frac{r_{ab}^i\,r_{ab}^j}{r_{ab}^2}\biggr) \nonumber
  \\ & +\frac1{4\,\pi\,r_{ab}^3} \bigl[ 2\,\bigl(\vec\sigma_a\cdot\vec
  r_{ab}\times\vec p_b - \vec\sigma_b\cdot\vec r_{ab}\times\vec p_a\bigr)
  \nonumber \\ & + \bigl(\vec\sigma_b\cdot\vec r_{ab}\times\vec p_b -
  \vec\sigma_a\cdot\vec r_{ab}\times\vec p_a\bigr)\bigr]\biggr\}\,.
\end{align}
We note that despite the presence of terms with $\ln Z$ in Eq.~(\ref{E5nlog}),
the correction $\cE^{(5)}_{\infty}({\rm nlog})$ does not have logarithmic
terms in its $1/Z$ expansion.

The recoil correction $\cE^{(5)}_M$ consists of four parts
\cite{pachucki:00:herec},
\begin{equation} \label{E51} \cE^{(5)}_M = \frac{m}{M}\bigl(
  \cE_{1}+\cE_{2}+\cE_{3}\bigr) + \langle H^{(5)}_{\rm fs, rec}\rangle \,,
\end{equation}
where
\begin{align} \label{E511} \cE_{1} = -3\, \cE^{(5)}_{\infty}+ \frac{4 Z}{3}
  \sum_a \langle \delta^3(r_a)\rangle - \frac{14}{3} \sum_{a<b} \langle
  \delta^3(r_{ab})\rangle \,,
\end{align}
\begin{align}
  \cE_{2} = &\ Z^2 \Biggl[ -\frac{2}{3}\ln (\Za)+\frac{62}{9} -\frac{8}{3}\ln
  \left(\frac{k_0}{Z^2}\right) \Biggr]\, \sum_a \langle \delta^3(r_a)\rangle
  \nonumber \\ & -\frac{14\, Z^2}{3}\,\sum_a \widetilde{Q}_a \,,
\end{align}
with $\widetilde{Q}_a$ defined analogously to Eq.~(\ref{Q}), and
$(m/M)\,\cE_{3}$ is the first-order perturbation of $\cE^{(5)}_{\infty}$ due to
the mass-polarization operator (\ref{MP}). The operator $H^{(5)}_{\rm fs,
  rec}$ yields a nonvanishing contribution to the fine-structure splitting
only. It is given by
\begin{equation}
  H^{(5)}_{\rm fs, rec} = \frac{m}{M}\,\frac{Z}{4\pi}\sum_{ab}
  \frac{\vec{r}_a}{r_a^3}\times \vec{p}_b \cdot\vec{\sigma}_a\,.
\end{equation}
We note that the last term in Eq.~(\ref{E511}) was omitted in the original
derivation of Ref.~\cite{pachucki:00:herec}.

The complete result for the $m\,\alpha^6$ correction $\cE^{(6)}_{\infty}$ to
the energy levels was derived by one of the authors (K.P.) in a series of
papers
\cite{pachucki:00:prl,pachucki:02:jpb,pachucki:06:hesinglet,pachucki:06:he}
\begin{align} \label{E6} \cE^{(6)}_{\infty} = &\ -\ln(\Za)\,\pi\,
\sum_{a<b}\langle
  \delta^3(r_{ab})\rangle+ E_{\rm sec} \nonumber \\& + \lbr H_{\rm
    nrad}^{(6)}+H_{R1}^{(6)}+H_{R2}^{(6)}+ H_{\rm fs}^{(6)}+ H_{\rm fs,
    amm}^{(6)} \rbr \,.
\end{align}
The first term in the above expression contains the complete logarithmic
dependence of the $m\,\alpha^6$ correction.  The part of it proportional to
$\ln\alpha$ was first obtained in Ref.~\cite{drake:93}.  The remaining
logarithmic part proportional to $\ln Z$ was implicitly present in formulas
reported in Ref.~\cite{pachucki:06:hesinglet,pachucki:06:he} (it originates
from the expectation value of the operator $1/r_{ab}^3$). In Eq.~(\ref{E6}),
we group all logarithmic terms together so that the remaining part does
not have any logarithms in its $1/Z$ expansion.

The term $E_{\rm sec}$ in Eq.~(\ref{E6}) is the second-order perturbation
correction induced by the Breit-Pauli Hamiltonian. (More specifically, it is
the finite residual after separating divergent contributions that cancel out
in the sum with the expectation value of the effective $m\alpha^6$
Hamiltonian.) The first part of the effective Hamiltonian, $H_{\rm
  nrad}^{(6)}$, originates from the non-radiative part of the electron-nucleus
and the electron-electron interaction. The next two terms, $H_{R1}^{(6)}$ and
$H_{R2}^{(6)}$, are due to the one-loop and two-loop radiative effects,
respectively. The last two parts $H_{\rm fs}^{(6)}$ and $H_{\rm fs,
  amm}^{(6)}$ are the spin-dependent operators first derived by Douglas and
Kroll \cite{douglas:74}. They do not contribute to the energies of the singlet
states and to the spin-orbit averaged levels. Expressions for these operators
are well known and are given, e.g., by Eqs.~(3) and (7) of
Ref.~\cite{pachucki:09:hefs}.  The non-radiative part of the $m\,\alpha^6$
effective Hamiltonian is rather complicated. For simplicity, we present it
specifically for a two-electron atom. The corresponding expression reads
\cite{pachucki:06:hesinglet,pachucki:06:he}
\begin{widetext}
  \begin{eqnarray} \label{H6nrad} H^{(6)}_{\rm nrad} &=& -\frac{\cE_0^3}{2}
    +\biggl[\biggl(-\cE_0 + \frac{3}{2}\,\vec p_2^{\;2}
    +\frac{1-2\,Z}{r_2}\biggr)\,\frac{Z\,\pi}{4}\,\delta^{3}(r_1)
    +(1\leftrightarrow 2)\biggr] \nonumber \\ && +\frac{\vec
      P^2}{6}\,\pi\,\delta^{3}(r)
    -\frac{(3+\vec\sigma_1\cdot\vec\sigma_2)}{24}\,\pi\,\vec
    p\,\delta^3(r)\,\vec p -\biggl(\frac{Z}{r_1} +
    \frac{Z}{r_2}\biggr)\,\frac{\pi}{2}\,\delta^{3}(r) \nonumber \\ &&
    +\biggl(\frac{13}{12}+\frac{8}{\pi^2}-\frac{3}{2}\,\ln(2)
    -\frac{39\,\zeta(3)}{4\,\pi^2}\biggr)\,\pi\,\delta^{3}(r)
    +\frac{\cE_0^2+2\,\cE^{(4)}}{4\,r} \nonumber \\ &&
    -\frac{\cE_0}{r^2}\,\frac{(31+5\,\vec\sigma_1\cdot\vec\sigma_2)}{32}
    -\frac{\cE_0}{2\,r}\,\bigg(\frac{Z}{r_1}+\frac{Z}{r_2}\biggr)
    +\frac{\cE_0}{4}\,\biggl(\frac{Z}{r_1}+\frac{Z}{r_2}\biggr)^2 \nonumber \\
    && -\frac{1}{r^2}\,\biggl(\frac{Z}{r_1}+\frac{Z}{r_2}-\frac{1}{r}\biggr)\,
    \frac{(23 + 5\,\vec\sigma_1\cdot\vec\sigma_2)}{32}
    -\frac{1}{4\,r}\,\biggl(\frac{Z}{r_1}+\frac{Z}{r_2}\biggr)^2 \nonumber \\
    &&
    +\frac{Z^2}{2\,r_1\,r_2}\,\biggl(\cE_0+\frac{Z}{r_1}+\frac{Z}{r_2}-\frac{1}{r}\biggr)
    -Z\,\biggl(\frac{\vec r_1}{r_1^3} - \frac{\vec
      r_2}{r_2^3}\biggr)\cdot\frac{\vec r}{r^3}\,
    \frac{(13+5\,\vec\sigma_1\cdot\vec\sigma_2)}{64} \nonumber \\ &&
    +\frac{Z}{4}\,\biggl(\frac{\vec r_1}{r_1^3} - \frac{\vec
      r_2}{r_2^3}\biggr)\cdot\frac{\vec r}{r^2}
    -\frac{Z^2}{8}\,\frac{r_1^i}{r_1^3}\,\frac{(r^i r^j -
      3\,\delta^{ij}\,r^2)}{r}\,\frac{r_2^j}{r_2^3} \nonumber \\ &&
    +\biggl[\frac{Z^2}{8}\,\frac{1}{r_1^2}\,\vec p_2^{\;2} +
    \frac{Z^2}{8}\,\vec p_1\,\frac{1}{r_1^2}\,\vec p_1 +\vec
    p_1\,\frac{1}{r^2}\,\vec p_1\,\frac{(47 +
      5\,\vec\sigma_1\cdot\vec\sigma_2)}{64} +(1\leftrightarrow 2)\biggr]
    \nonumber \\ &&
    +\frac{1}{4}\,p_1^i\,\biggl(\frac{Z}{r_1}+\frac{Z}{r_2}\biggr)\,
    \frac{(r^i\,r^j + \delta^{ij}\, r^2)}{r^3}\, p_2^j
    +P^i\,\frac{(3\,r^i\,r^j - \delta^{ij}
      r^2)}{r^5}\,P^j\,\frac{(-3+\vec\sigma_1\cdot\vec\sigma_2)}{192}
    \nonumber \\ &&
    -\biggl[\frac{Z}{8}\,p_2^k\,\frac{r_1^i}{r_1^3}\,\biggl(\delta^{jk}\,\frac{r^i}{r}
    - \delta^{ik}\, \frac{r^j}{r} - \delta^{ij}\, \frac{r^k}{r} - \frac{r^i\,
      r^j\, r^k}{r^3}\biggr)\,p_2^j +(1\leftrightarrow 2)\biggr] \nonumber \\
    && -\frac{\cE_0}{8}\,p_1^2\,p_2^2-\frac{1}{4}\,p_1^2\,\biggl(\frac{Z}{r_1}
    +\frac{Z}{r_2}\biggr)\,p_2^2 +\frac{1}{4}\,\vec p_1\times\vec
    p_2\,\frac{1}{r}\,\vec p_1\times\vec p_2 \nonumber \\ &&
    +\frac{1}{8}\,p_1^k\,p_2^l\,\biggl(-\delta^{jl}\,\frac{r^i\,r^k}{r^3} -
    \delta^{ik}\,\frac{r^j\,r^l}{r^3} + 3\,\frac{r^i\,r^j\,r^k\,r^l}{r^5}
    \biggr)\, p_1^i\,p_2^j +\ln(Z)\,\pi\,\delta^3(r)\,,
  \end{eqnarray}
\end{widetext}
where $\vec P = \vec p_1 + \vec p_2$, $\vec p = (\vec p_1-\vec p_2)/2$, $\vec
r = \vec r_1-\vec r_2$. We note that the operator $H^{(6)}_{\rm nrad}$ is
defined in such a way that its expectation values does not contain any
logarithmic terms in the $1/Z$ expansion, as the last term of
Eq.~(\ref{H6nrad}) is compensated by the corresponding contribution from the
$1/r^3$ operator.  

The effective Hamiltonians induced by the radiative effects
are \cite{korobov:01,yelkhovsky:01,pachucki:06:hesinglet}
\begin{align}
  H^{(6)}_{R1} &\ = Z^2\,\biggl[\frac{427}{96}-2\,\ln(2)\biggr]\,\pi\,
  \bigl[\delta^3(r_1)+\delta^3(r_2)\bigr] \nonumber \\ & +\biggl[
  \frac{6\,\zeta(3)}{\pi^2}-\frac{697}{27\,\pi^2}-8\,\ln(2)+\frac{1099}{72}
  \biggr]\,\pi\,\delta^3(r),
\end{align}
and
\begin{align}
  H^{(6)}_{R2} &\ =
  Z\,\biggl[-\frac{9\,\zeta(3)}{4\,\pi^2}-\frac{2179}{648\,\pi^2}+
  \frac{3\,\ln(2)}{2}-\frac{10}{27}\biggr]\,\pi\, \nonumber \\ & \times
  \bigl[\delta^3(r_1)+\delta^3(r_2)\bigr]
  +\biggl[\frac{15\,\zeta(3)}{2\,\pi^2}+\frac{631}{54\,\pi^2} \nonumber \\ &
  -5\,\ln(2)+\frac{29}{27} \biggr]\,\pi\,\delta^3(r)\,.
\end{align}

The second-order correction can be represented as
\begin{align} \label{sec6} E_{\rm sec} &\ = \lbr H^{(4)^{\prime}}_{\rm nfs}
  \frac1{(E_0-H_0)'} H^{(4)^{\prime}}_{\rm nfs} \rbr \nonumber \\ & + 2\,\lbr
  H^{(4)}_{\rm nfs} \frac1{(E_0-H_0)'} H^{(4)}_{\rm fs} \rbr \nonumber \\ & +
  \lbr H^{(4)}_{\rm fs} \frac1{(E_0-H_0)'} H^{(4)}_{\rm fs} \rbr\,,
\end{align}
where $H^{(4)}_{\rm nfs}$ and $H^{(4)}_{\rm fs}$ are the spin-independent and
spin-dependent parts of the Breit-Pauli Hamiltonian (\ref{H4}), respectively.
The operator $H^{(4)^{\prime}}_{\rm nfs}$ is obtained from $H^{(4)}_{\rm nfs}$
by a transformation that eliminates divergences in the second-order matrix
elements \cite{pachucki:06:hesinglet}. The transformed operator is given by
\begin{align}
  H^{(4)^{\prime}}_{\rm nfs} & = -\frac{1}{2}\,(\cE_0-V)^2
  -p_1^i\,\frac{1}{2\,r}\,\biggl(\delta^{ij}+\frac{r^i\,r^j}{r^2}\biggr)\,p_2^j
  \nonumber \\ & +\frac{1}{4}\,\vec \nabla_1^2 \, \vec \nabla_2^2
  -\frac{Z}{4}\,\frac{\vec r_1}{r_1^3}\cdot\vec \nabla_1
  -\frac{Z}{4}\,\frac{\vec r_2}{r_2^3}\cdot\vec \nabla_2 \,,
\end{align}
where $\vec \nabla_1^2 \, \vec \nabla_2^2$ is understood as a differentiation
of the wave function on the right hand side as a function (omitting
$\delta^3(r)$) and $V = -Z/r_1-Z/r_2+1/r$.

An intriguing feature of the formulas presented in this section is that the
logarithmic dependence of all of them enters only in the form of $\ln
(\Za)$. This is not at all obvious {\em a priori} 
since $\ln (\Za)$ appears naturally
only in contributions induced by the electron-nucleus interaction. The
effects of the electron-electron interaction usually yield logarithms of $\alpha$,
whereas logarithms of $Z$ are implicitly present in matrix elements of
singular operators. The fact that logarithms of $\alpha$ and
logarithms of $Z$ have the coefficients that match each other comes
``accidentally'' from the derivation.

The complete result for the corrections of order $m\,\alpha^7$ for the helium
Lamb shift is not presently available (it is known for the fine-structure
splitting only \cite{pachucki:06:prl:he,pachucki:09:hefs}).  One can, however,
easily generalize some of the hydrogenic results, namely those that are
proportional to the electron density at the nucleus. These are (i) the
one-loop radiative correction of order $m\alpha\,(\Za)^6\,\ln^2(\Za)^{-2}$,
(ii) the two-loop radiative correction of order $m\alpha^2\,(\Za)^5$, and
(iii) the nonrelativistic correction due to the finite nuclear size. The first
two effects yield the main contribution to the higher-order remainder
function of $S$ states in light hydrogen-like ions. We expect that they dominate for
light helium-like ions as well. 

Following Ref.~\cite{drake:88:cjp}, we approximate the higher-order radiative
(``rad'') and the finite-nuclear-size (``fs'') correction to the energies of
helium-like ions by
\begin{align}\label{ho}
  \cE^{(7+)}_{\rm rad} =&\ \cE^{(7+)}_{{\rm rad}, H} \, \frac{\bigl< \sum_i
    \delta^3(r_i)\bigr>}{\bigl< \sum_i \delta^3(r_i)\bigr>_H}\,,
  \\
  E_{\rm fs} =&\ E_{{\rm fs}, H} \, \frac{\bigl< \sum_i
    \delta^3(r_i)\bigr>}{\bigl< \sum_i \delta^3(r_i)\bigr>_H}\,,
  \label{hoa}
\end{align}
where the subscript $H$ corresponds to the ``hydrogenic'' limit, i.e., the
limit of the non-interacting electrons, and
\begin{align}
  \bigl< \sum_i \delta^3(r_i)\bigr>_{H} = \frac{Z^3}{\pi}\left( 1+
    \frac{\delta_{l,0}}{n^3}\right)\,.
\end{align}
The approximation of Eqs.~(\ref{ho}) and (\ref{hoa}) is exact for the
above-mentioned corrections proportional to the electron density at the
nucleus.  It is expected also to provide a meaningful estimate for
contributions that weakly depend on $n$ (such as the nonlogarithmic radiative
correction of order $m\alpha\,(\Za)^6$). Moreover, this approximation is exact
to the leading order in the $1/Z$ expansion, thus providing a meaningful
estimate for high-$Z$ helium-like ions as well.

For all the states under consideration except $2^1P_1$ and $2^3P_1$, the
``hydrogenic'' remainder function is just the sum
of the corresponding remainders for the two electrons in the configuration,
\begin{align}
  \cE^{(7+)}_{{\rm rad}, H} = \cE^{(7+)}_{\rm rad}(1s) + \cE^{(7+)}_{\rm
    rad}(nlj)\,.
\end{align}
For the $2^1P_1$ and $2^3P_1$ states, the Dirac levels need to be first
transformed from the $jj$ to the $LS$ coupling and thus \cite{drake:88:cjp}
\begin{align} \label{pho} \cE^{(7+)}_{{\rm rad}, H}(2^1P_1) &= \cE^{(7+)}_{\rm
    rad}(1s) + \frac23\,\cE^{(7+)}_{\rm rad}(2p_{3/2}) +
  \frac13\,\cE^{(7+)}_{\rm rad}(2p_{1/2})\,,
  \nonumber \\
  \cE^{(7+)}_{{\rm rad}, H}(2^3P_1) &= \cE^{(7+)}_{\rm rad}(1s) +
  \frac13\,\cE^{(7+)}_{\rm rad}(2p_{3/2}) + \frac23\,\cE^{(7+)}_{\rm
    rad}(2p_{1/2})\,.
\end{align}

In our calculation, the one-electron remainder function $\cE^{(7+)}_{\rm
  rad}(nlj)$ includes all known contributions of order $m\alpha^7$ and higher
coming from (i) the
one-loop radiative correction, (ii) the two-loop radiative correction, (iii)
the three-loop radiative correction, see the review in Ref.~\cite{mohr:08:rmp}
and Ref.~\cite{yerokhin:09:sese} for an update on the two-loop remainder
function.

Besides the finite-nuclear-size and radiative corrections, 
there are also non-radiative
effects, denoted as $\cE^{(7+)}_{\rm nrad}$ and 
estimated within the $1/Z$ expansion. More specifically, we include 
the higher-order remainder due to the one-electron Dirac energy and due
to the one-photon exchange correction. They enter at the order $m \alpha^8$
only but are enhanced by factors of $Z^8$ and $Z^7$, respectively. Despite this
enhancement, numerical contributions of these effects are rather small for the ions
considered in the present work.


\section{Results and discussion}


\subsection{Numerical results}

The nonrelativistic energies and wave functions are obtained by minimizing the
energy functional with the basis set constructed with the fully correlated
exponential functions. The choice of the basis set and the general strategy of
optimization of the nonlinear parameters follow the main lines of the
approach developed 
by Korobov \cite{korobov:00,korobov:02}. The calculational scheme is described in
previous publications
\cite{pachucki:06:hesinglet,pachucki:06:he,pachucki:09:hefs} and will not be
repeated here.  Numerical values of the nonrelativistic energies of
helium-like ions with the nuclear charge $Z=3\ldots 12$ are presented in
Table~\ref{tab:E2}. The results were obtained with $N=2000$ basis functions
and are accurate to about 18 decimals (more than shown in the table). The
energy levels of the helium atom traditionally attract special attention, so we
present the corresponding results in full length below.  Our numerical values
of the upper variational limit of the nonrelativistic energies of helium are
\begin{align} \label{E2Hea}
  \cE^{(2)}_{\infty}(1^1S) &\ =  -2.903\,724\,377\,034\,119\,598\,310\,^{+0}_{-2} \,,\\
  \cE^{(2)}_{\infty}(2^1S) &\ =  -2.145\,974\,046\,054\,417\,415\,799\,^{+0}_{-8} \,,\\
  \cE^{(2)}_{\infty}(2^3S) &\ =  -2.175\,229\,378\,236\,791\,305\,738\,977\,^{+0}_{-2} \,,\\
  \cE^{(2)}_{\infty}(2^1P) &\ =  -2.123\,843\,086\,498\,101\,359\,246\,^{+0}_{-2} \,,\\
  \cE^{(2)}_{\infty}(2^3P) &\ =
  -2.133\,164\,190\,779\,283\,205\,146\,96\,^{+0}_{-10} \,.
  \label{E2Heb}
\end{align}
The value for the ground state is given only for completeness, since much more
accurate numerical results are available in the literature
\cite{korobov:02,schwartz:06}.  The numerical results for the leading
relativistic correction $\cE^{(4)}$ are summarized in Table~\ref{tab:E4}.

The leading QED correction $\cE^{(5)}$ is given by Eqs.~(\ref{E5log}),
(\ref{E5nlog}), and (\ref{E51}). Computationally the most problematic part of
it is represented by the Bethe logarithm $\ln (k_0)$ and its mass-polarization
correction $\ln (k_0)_M$. Accurate calculations of $\ln (k_0)$ were performed
by Drake and Goldman \cite{drake:99:cjp} for helium-like like atoms with $Z\le
6$ and by Korobov \cite{korobov:04} for $Z=2$. Calculations of the recoil
correction to the Bethe logarithm were reported by Pachucki and Sapirstein
\cite{pachucki:00:herec} for $Z=2$ and by Drake and Goldman
\cite{drake:99:cjp} for $Z\le 6$. In the present investigation, we perform
accurate evaluations of the Bethe logarithm $\ln (k_0)$ and its recoil
correction $\ln (k_0)_M$ for helium-like ions with $Z\le 12$. The
calculational approach is described in Appendix~\ref{app:bethe}.

Table~\ref{tab:bethelog} summarizes the numerical results obtained and
gives a comparison with the previous calculations.
Numerical values for the Bethe logarithm are presented for the difference
$\ln(k_0)-\ln (Z^2)$ since this difference has a weak $Z$-dependence and does
not contain any logarithms in its $1/Z$ expansion. 
The table also lists the coefficients of the $1/Z$ expansion of 
$\ln(k_0/Z^2)$. 
The leading coefficient $c_0$ is known from the hydrogen theory; accurate numerical
values can be found in Ref.~\cite{drake:90}. 
The  higher-order coefficients were
obtained by fitting our numerical data. It is interesting to compare them with
the analogous results reported previously by Drake and Goldman \cite{drake:99:cjp}. 
For the next-to-the-leading coefficient $c_1$, the results agree
up to about 4-5 digits for $S$ states and about 3-4 digits for $P$ states. For the
higher-order coefficients, the agreement gradually deteriorates. However, the
results for the sum of the two expansions agree very well with each other. More
specifically, the maximal absolute deviation between
the values of the Bethe logarithms for $Z > 12$
obtained with our $1/Z$-expansion coefficients and with those by
Drake and Goldman is $1\times10^{-8}$
for the $1^1S$ state, $3\times10^{-8}$ for the $2^1S$ state, $6\times10^{-9}$
for the $2^3S$ state, $1\times10^{-7}$ for the $2^1P$ state, and $6\times10^{-8}$
for the $2^3P$ state. So, the accuracy of 
these expansions is sufficient for most practical purposes.

Another part of the calculation of $\cE^{(5)}$ that needs a separate
discussion is the evaluation of the expectation value of the singular operator
$1/r^3$, which is defined by Eq.~(\ref{rcube}). The calculational approach
is described in Appendix~\ref{app:rcube}.  Total results for the
logarithmic and the nonlogarithmic part of the leading QED correction are
summarized in Tables~\ref{tab:E5log} and \ref{tab:E5nlog}, respectively. The
results are in good agreement with the previous calculations
\cite{drake:88:cjp}.

Table~\ref{tab:E6} presents the numerical values of the $m\alpha^6$
correction, the main result of this investigation. The corresponding
calculations for atomic helium were reported in
Refs.~\cite{pachucki:06:hesinglet,pachucki:06:he}; our present numerical
values agree with the ones obtained previously.  Calculations performed in
this work for helium-like ions were accomplished along the lines described in
Refs.~\cite{pachucki:06:hesinglet,pachucki:06:he}. Here we only note that
calculations for higher values of $Z$ often exhibit a slower numerical
convergence (and numerical stability) than for helium, especially so for the
second-order corrections involving singular operators. The variational
optimization of nonlinear parameters for the symmetric second-order
corrections was performed in several steps with increasing the number of basis
functions on each step up to $N=1000$ or $1200$. The final values were
obtained with merging several basis sets and enlarging the number of
functions up to $N=5000-7000$.  The nonsymmetric second-order corrections were
evaluated as described in Ref.~\cite{pachucki:09:hefs}. The calculations were
performed in the quadruple, sixtuple, and octuple arithmetics implemented in
Fortran~95 by V.~Korobov \cite{korobov:priv}.

Table~\ref{tab:E7} presents the results for the finite nuclear size correction
and approximate values of the higher-order
($m\,\alpha^7$ and higher) correction to the {\em ionization energy}.  The
uncertainty of the total theoretical prediction originates from the
higher-order radiative effects; it was estimated by dividing the absolute
value of this correction by $Z$. The values for the
root-mean-square radius of nuclei were taken from Ref.~\cite{angeli:04}.


\subsection{Comparison with the all-order approach}

In this subsection we discuss the calculational results obtained for the
$m\,\alpha^6$ correction in more detail and make a comparison with the results
obtained previously within the all-order, $1/Z$-expansion approach.  
The logarithmic part of the
correction, $\cE^{(6)}({\rm log})$, behaves as $m\,\alpha^3(\Za)^3$ for large
$Z$ and thus corresponds to diagrams with three photon exchanges that have not
yet been addressed within the all-order approach.  The nonlogarithmic part
$\cE^{(6)}({\rm nlog})$, however, contains some pieces that are known and 
identified below.

For all states except $2^1P_1$ and $2^3P_1$, the leading term of the $1/Z$
expansion of $\cE^{(6)}({\rm nlog})$ is of order $m\,(\Za)^6$ and comes from
the $\Za$ expansion of the one-electron Dirac energy. For the $2^1P_1$ and
$2^3P_1$ states, the leading term is of the previous order in $1/Z$,
$m\,(\Za)^6 Z$, and is due to the mixing of these levels.  More specifically, the
mixing contribution is $ \delta E_{\rm mix} = \left|\langle
  2^1P_1|H^{(4)}|2^3P_1\rangle\right|^2/[E_0(2^1P_1)-E_0(2^3P_1)] $ for the
$2^1P_1$ state and that with the opposite sign, for $2^3P_1$. The contribution
of order $m\,(\Za)^6$ for the mixing states comes from the expansion of the
one-electron Dirac energy and from the expansion of $\delta E_{\rm mix}$.

The next term of the $1/Z$ expansion is of order $m\,\alpha (\Za)^5$ and comes
from the one-electron one-loop radiative correction and from the one-photon
exchange correction. The radiative part is well known \cite{mohr:08:rmp}. The
part due to the one-photon exchange was obtained for the $1^1S$, $2^3S$,
$2^3P_0$, and $2^3P_2$ states analytically in Ref.~\cite{mohr:85:pra} and for
the other states numerically in this work. For the $2^1P_1$ and $2^3P_1$
states, there is a small additional mixing contribution, which we were unable
to determine unambiguously.

The first two coefficients of the $1/Z$ expansion of $\cE^{(6)}({\rm nlog})$
are listed in Table~\ref{tab:E6}. A fit of our numerical data agrees well with
these coefficients. The agreement observed shows consistency of our numerical
results with independent calculations at the level of the one-photon effects.
We now turn to the contribution of order $m\,\alpha^2(\Za)^4$. This
contribution is induced by nontrivial two-photon effects, so that a
comparison drawn for this part
will yield a much more stringent test of consistency of different
approaches.

The part of $\cE^{(6)}({\rm nlog})$ that is of order $m\,\alpha^2 (\Za)^4$ is
implicitly present in the two-electron QED contribution calculated numerically
in Ref.~\cite{artemyev:05:pra} to all orders in $\Za$.  This contribution can
be represented as (see Eq.~(72) of Ref.~\cite{artemyev:05:pra})
\begin{align}\label{2elQED}
  \Delta E^{\rm QED}_{\rm 2el} = m\,\alpha^2\,(\Za)^3\,\Bigl[ a_{31}\,\ln
  (\Za)^{-2}+ a_{30}+ (\Za)\,G(Z)\Bigr]\,,
\end{align}
where the remainder function $G(Z)$ incorporates all higher orders in $\Za$.
The two-electron QED correction comprises the so-called screened self-energy
and vacuum-polarization contributions and the part of the two-photon exchange
correction that is beyond the Breit approximation.

The coefficients $a_{31}$ and $a_{30}$ in Eq.~(\ref{2elQED}) correspond to the
second term of the $1/Z$ expansion of the leading QED correction
$\cE^{(5)}_{\infty}$. More specifically, $a_{31}$ corresponds to the
coefficient $c_1$ from Table~\ref{tab:E5log} and $a_{30}$, to that from
Table~\ref{tab:E5nlog}. The $Z=0$ limit of the higher-order remainder function
$G(Z)$ corresponds to the third coefficient of the $1/Z$ expansion of the
correction $\cE^{(6)}({\rm nlog})$, $G(0) = c_2$, for all states except
$2^1P_1$ and $2^3P_1$. The values of $c_2$ obtained by fitting our numerical
data are listed in Table~\ref{tab:E6}. For the $2^1P_1$ and $2^3P_1$ states,
the coefficient $c_2$ is not directly comparable with the all-order results
because of the mixing effects.

\begin{figure*}[thb]
  \centerline{\includegraphics[width=\textwidth]{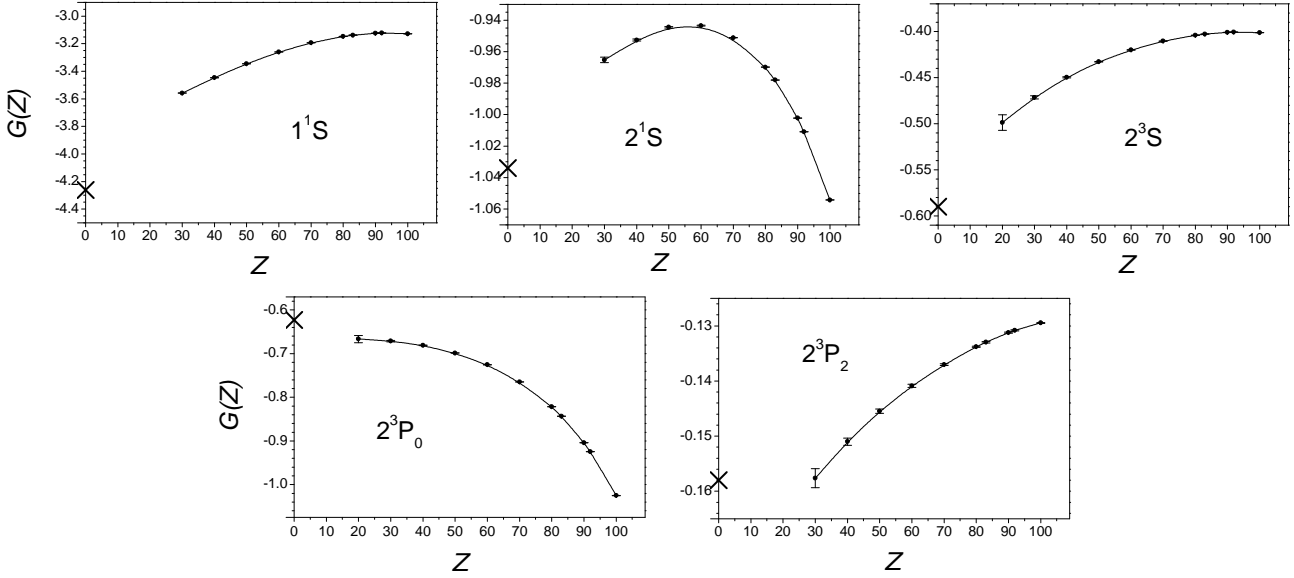}}
\caption{The higher-order remainder function $G(Z)$ from Eq. (\ref{2elQED}) 
  inferred from the all-order
  numerical results of Ref.~\cite{artemyev:05:pra} for the two-electron QED
  correction, in comparison with the $Z=0$ limit obtained by fitting the $1/Z$
  expansion of the $m\alpha^6$ correction calculated in this work (denoted by
  the cross on the $y$ axis). The all-order results for $Z$ smaller than 30
  (in some cases, 20) were left out since their numerical accuracy turns out
  to be not high enough.
  \label{fig:1} }
\end{figure*}

The higher-order remainder function $G(Z)$ inferred from the numerical results
of Ref.~\cite{artemyev:05:pra} is plotted in Fig.~\ref{fig:1}, together with
its limiting value at $Z=0$ obtained by a fit of our numerical data.  It should
be stressed that the identification of the remainder implies a great deal of
numerical cancellations, especially for the all-order results. The comparison
drawn in Fig.~\ref{fig:1} provides a stringent cross-check of the two
complementary approaches. The visual agreement between the results is very
good for the $S$ states, whereas for the $P$ states a slight disagreement
seems to be present.

It is tempting to merge the all-order and the $\Za$-expansion results by
fitting the all-order data for $G(Z)$ towards
lower values of $Z$. However, we
do not attempt to do this in the present work. The reasons are, first, that
the numerical accuracy of the all-order results is apparently not high enough
and, second, that the expansion of the remainder function $G(Z)$ contains
terms $(\Za)\ln^2(\Za)$ and $(\Za)\ln (\Za)$, which cannot be reasonably
fitted with numerical data available in the high-$Z$ region only.


\subsection{Total energies}

Our total results for the ionization energy of the $n=1$ and $n=2$
states of helium-like atoms with the nuclear charge $Z=2\ldots 12$ are listed
in Table~\ref{tab:total}. The following values of fundamental constants were
employed \cite{mohr:08:rmp}, $R_{\infty} = 10\,973\,731.568\,527(73)\ {\rm
  m}^{-1}$ and $\alpha^{-1} = 137.0359\,999\,679(94)$. The atomic masses were
taken from Ref.~\cite{audi:03}.

The results for atomic helium presented in Table~\ref{tab:total} differ from
those reported previously only because of the different approximate treatment of
the higher-order ($m\,\alpha^7$ and higher) contribution employed in this
work.  For the $S$ states of helium, the present values are practically
equivalent to those of Refs.~\cite{pachucki:06:hesinglet,pachucki:06:he}. (The
difference is that now we include some contributions of order $m\,\alpha^8$
and higher, which are negligible for helium but become noticeable for
higher-$Z$ ions.) However, for the helium $P$ states, our present estimate of
the higher-order contribution is by about 1~MHz higher than that of
Ref.~\cite{pachucki:06:he}. The reason is that the one-electron radiative
correction of the $p$ electron state was previously not included into the
approximation (\ref{ho}). It is included now [see Eq.~(\ref{pho})] in order to
recover the correct asymptotic behaviour of the radiative correction in the
high-$Z$ limit.

A selection of our theoretical results for transition energies is compared
with the theory by Drake \cite{drake:88:cjp,drake:05:springer} and with
experimental data in Table~\ref{tab:transen}. Agreement between theory and
experiment is excellent in all cases studied. We
observe a distinct improvement in theoretical accuracy as compared to the
previous results by Drake. This improvement is due to the complete treatment
of the corrections of order $m\alpha^6$ and $m^2/M\alpha^5$ accomplished in
this work.

Theoretical results for the fine structure splitting intervals $2^3P_0-2^3P_1$
and $2^3P_1-2^3P_2$ are not analyzed in the present work. This is because
these intervals can nowdays be calculated more accurately (complete up to
order $m\alpha^7$), like it was recently done for helium
\cite{pachucki:09:hefs}. We intend to perform such a calculation in a
subsequent investigation.

Among the results listed in Table~\ref{tab:total} for helium-like ions, the
ground-state energy of the carbon ion is of particular importance, because it
is used in the procedure of the adjustment of fundamental constants
\cite{mohr:05:rmp} for the determination of the mass of $^{12}$C$^{4+}$ and,
consequently, of the proton mass from the Penning trap measurement by Van Dyck
{\em et al.} \cite{vandyck:99}.  Our result for the ground-state ionization
energy of helium-like carbon is
\begin{align}
  E(^{12}{\rm C}^{4+}) = -3\,162\,423.60(32) \ {\rm cm}^{-1}\,,
\end{align}
which is in agreement with the previous result by Drake \cite{drake:88:cjp} of
$-3\,162\,423.34(15)\ {\rm cm}^{-1}$. We note that despite the fact that our
calculation is by an additional order of $\alpha$ more complete than that by
Drake, our estimate of uncertainty is more conservative.

In summary, significant progress has been achieved during last decades both in
experimental technique and theoretical calculations of helium-like atoms.  In
the present investigation, we performed a calculation of the energy levels of
the $n=1$ and $n=2$ states of light helium-like atoms with the nuclear charge
$Z=2\ldots 12$, within the approach complete up to orders $m\alpha^6$ and
$m^2/M\alpha^5$. An extensive analysis of the $1/Z$ expansion of individual
corrections was carried out and comparison with results of the complementary
approach was made whenever possible. Our general conclusion is that the results
obtained within the approaches based on the $\Za$ and the $1/Z$ expansion are
consistent with each other up to a high level of precision. However, further
improvement of numerical accuracy of the all-order, $1/Z$-expansion results
and their extension into the lower-$Z$ region is needed in order to safely
merge the two complementary approaches.


\section*{Acknowledgments}
We wish to thank Tomasz Komorek for participation in the beginning of
the project.  Computations reported in this work were performed on the
computer clusters of St.~Petersburg State Polytechnical University and of
Institute of Theoretical Physics, University of Warsaw.  The research was
supported by NIST through Precision Measurement Grant PMG 60NANB7D6153.
V.A.Y.  acknowledges additional support from the ``Dynasty'' foundation and
from RFBR (grant No.~10-02-00150-a).

%
%
%
\appendix

\section{Bethe logarithm}
\label{app:bethe}

Following the approach of Refs.~\cite{schwartz:61,korobov:99}, 
the Bethe logarithm (\ref{bethe}) is expressed in terms of an integral
over the momentum of the virtual photon, 
\begin{align} \label{be1} \ln (k_0) = \frac1D \lim_{K\to \infty} \Biggl[
  \bigl< {\vec{\nabla}}^2\bigr>\,K + D\, \ln (2K) + \int_0^K dk\, k J(k)
  \Biggr]\,,
\end{align}
where $D = 2\pi Z \bigl< \delta^3(r_1)+\delta^3(r_2)\bigr>$, $\vec{\nabla}
\equiv \vec{\nabla}_1+\vec{\nabla}_2$, and
\begin{align}
  J(k) = \biggl< \vec{\nabla}\, \frac1{E_0-H_0-k}\, \vec{\nabla} \biggr>\,.
\end{align}
For the purpose of numerical evaluation, the integration over the photon
momentum $k$ is divided into two regions by introducing the auxiliar parameter
$\kappa$,
\begin{align} \label{be2} \ln (k_0) = R(\kappa) + \frac1{D}
  \int_0^{\kappa}dk\, k\, J(k) + \int_{\kappa}^{\infty} dk\,
  \frac{w(k)}{k^2}\,,
\end{align}
where the function $w(k)$ represents the residual obtained from $J(k)$ by
removing all known terms of the large-$k$ asymptotics,
\begin{align} \label{be3} w(k) = \frac{k^3}{D}J(k) + \frac{k^2}{D}\bigl<
  {\vec{\nabla}}^2\bigr> + k - 2\sqrt{2}\,Z k^{1/2}+ 2Z^2 \ln k\,,
\end{align}
and $R(\kappa)$ is a simple function obtained by integrating out the separated
asymptotic expansion terms,
\begin{align}
  R(\kappa) = \kappa \frac{\bigl< {\vec{\nabla}}^2\bigr>}{D} + \ln (2\kappa) +
  \frac{4\sqrt{2}Z}{\kappa^{1/2}} -\frac{2Z^2(\ln \kappa+1)}{\kappa}\,.
\end{align}

The calculational scheme employed for the evaluation of Eq.~(\ref{be2}) is
similar to that previously used \cite{pachucki:09:hefs} for the relativistic
corrections to the Bethe logarithm. At the first step, a careful optimization
of nonlinear basis-set parameters was carried out for a sequence of scales of
the photon momentum: $k_i = 10^i$ and $i = 1,\ldots,i_{\rm max}$, with $i_{\rm
  max}=5$ for the $S$ states and $i_{\rm max}=4$ for the $P$ states.  The
optimization was performed with incrementing the size of the basis until the
prescribed level of convergence is achieved for the function $w(k)$. At the
second step, the integrations of the photon momentum $k$ were performed. For a
given value of $k$, the function $J(k)$ was calculated with a basis obtained
by merging together the optimized bases for the two closest $k_i$ points, thus
essentially doubling the number of the basis functions. The function $w(k)$
was obtained from $J(k)$ according to Eq.~(\ref{be3}).

The integral over $k\in[0,\kappa]$ was calculated analytically, by
diagonalizing the Hamiltonian matrix and using the spectral representation of
the propagator. The value of the auxiliar parameter $\kappa$ was set to
$\kappa=100$. The integral over $k\in[\kappa,\infty)$ was separated into two
parts, $k< 10^{i_{\rm max}}$ and $k> 10^{i_{\rm max}}$. The first part was
evaluated numerically by using the Gauss-Legendre quadratures, after the
change of variables $t = 1/k^2$. The second part was obtained by integrating
the asymptotic expansion of the function $w(k)$. The coefficients of this
expansion were obtained by fitting the numerical data for $w(k)$ to the form
\begin{eqnarray} \label{fit} w(k) = {\rm pol}\left(\frac1{\sqrt{k}}\right) +
  \frac{\ln k}{k}\,{\rm pol}\left(\frac1k\right)\,,
\end{eqnarray}
where ${\rm pol}(x)$ denotes a polynomial of $x$. The total number of fitting
parameters was about $9-11$. The range of $k$ to be fitted and the exact form
of the fitting function were optimized so as to yield the best possible
results for the known asymptotic expansion terms of $J(k)$.

The first-order perturbation of the Bethe logarithm by the mass-polarization
operator can be represented as \cite{pachucki:00:herec}
\begin{align}\label{be4}
  \ln (k_0)_M = &\ \frac{m}{M}\Biggl[R_M(\kappa) + \frac1{D}
  \int_0^{\kappa}dk\, k\, J_M(k) \nonumber \\ & + \int_{\kappa}^{\infty} dk\,
  \frac{w_M(k)}{k^2}\Biggr]\,,
\end{align}
where
\begin{align}
  & J_M(k) = \ 2\,\biggl<\phi \left| \vec{\nabla}\, \frac1{E_0-H_0-k}\,
    \vec{\nabla}\right| \delta \phi \biggr>
  \nonumber \\
  & + \biggl<\phi \biggl| \vec{\nabla}\, \frac1{E_0-H_0-k}\, \bigl[ \delta E -
  \vec{p}_1\cdot\vec{p}_2 \bigr]
  \frac1{E_0-H_0-k}\, \vec{\nabla}\biggr| \phi \biggr>\,,
\end{align}
and $\delta E = \bigl< \vec{p}_1\cdot \vec{p_2}\bigr>$. The perturbed wave
function $\delta \phi$ is defined by
\begin{align}
  |\delta \phi\bigr> = |\delta_M \phi\bigr> -|\phi\bigr> \frac{\delta_M
    D}{D}\,,
\end{align}
where $\delta_M$ stands for the first-order perturbation induced by the
operator $\vec{p}_1\cdot\vec{p}_2$. The asymptotic expansion of $J_M(k)$ is
much simpler than that of $J(k)$ and $w_M(k)$ is just
\begin{align}
  w_M(k) = \frac{k^3}{D}J_M(k) + \frac{k^2}{D}2\bigl<\phi|
  {\vec{\nabla}}^2|\delta \phi \bigr> \,.
\end{align}
Correspondingly, the function $R_M(\kappa)$ is
\begin{align}
  R_M(\kappa) = \frac{2\kappa}{D} \bigl<\phi| {\vec{\nabla}}^2|\delta \phi
  \bigr>\,.
\end{align}
The numerical evaluation of Eq.~(\ref{be4}) was performed in a way similar to
that for the plain Bethe logarithm. In particular, the same sets of optimized
nonlinear parameters were used. Since a high accuracy is not needed for this
correction, a somewhat simplified calculational scheme was used in this case.
The high-energy part of the photon-momentum integral, $k\in [100,\infty)$, was
evaluated by integrating the fitted asymptotic expansion for $w_M(k)$, which
was taken to be of the form (\ref{fit}) with $6-9$ fitting parameters.


\section{Expectation value of $\bm{1}/\bm{r}^{\bm{3}}$}
\label{app:rcube}

The definition of the expectation value of the regularized operator $1/r^3$ is
given by Eq.~(\ref{rcube}). With the basis-set representation of the wave
function employed in this work, a typical singular integral to be calculated is
\begin{align}
  I_\epsilon = \frac1{16\pi^2} \int d^3r_1\, d^3r_2\, \frac{\exp (-\alpha r_1
    -\beta r_2 -\gamma r)}{r^3}\,\Theta(r-\epsilon)\,.
\end{align}
The straightforward way is to evaluate this integral analytically for a finite value of the
regulator $\epsilon$ and then expand it in small $\epsilon$. This way is possible,
but we prefer to use a simpler procedure, which is also the closest to the method of
evaluation of the regular integrals. 

We recall that all regular integrals are immediately obtained from the master
integral
\begin{align}
  \frac1{16\pi^2} \int d^3r_1\, d^3r_2\, & \frac{\exp (-\alpha r_1 -\beta r_2
    -\gamma r)}{r_1r_2r} \nonumber \\ & =
  \frac1{(\alpha+\beta)(\beta+\gamma)(\alpha+\gamma)}
\end{align}
by formal differentiation or integration with respect to the corresponding
parameters. The differentiation over $\alpha$ and $\beta$ and an
integration over $\gamma$ delivers a result for the integral of the type
$1/r^2$. This integral is convergent, so the result is exact. The second
integration over $\gamma$ (which would yield an integral of the type $1/r^3$)
is divergent. The simplest way to proceed is as follows. We 
introduce a cutoff parameter for large values of $\gamma$, evaluate the integral over
$\gamma$, and drop all cutoff-dependent terms. The expression obtained in this way
differs from the correct one by a $\gamma$-independent constant only,
which can be proved by differentiating with respect to $\gamma$.

The missing constant is most easily recovered by considering the behavior of
the integral $I$ when $\gamma\to\infty$. For very large $\gamma$, only the
region of very small $r$ contributes and we have
\begin{align}
  I_{\epsilon} &= 2\int_{\epsilon}^{\infty}dr\,r \int_0^{\infty}dr_1\,r_1
  \int_{|r_1-r|}^{r_1+r}dr_2\,r_2\, \frac{e^{-\alpha r_1 -\beta r_2 -\gamma
      r}}{r^3} \nonumber \\ & \approx \frac{2}{(\alpha+\beta)^3}\,
  \int_{\epsilon}^{\infty} dr \frac{e^{-\gamma r}}{r} \,.
\end{align}
Therefore,
\begin{align}
  I_{\rm reg}\equiv \lim_{\epsilon\to0}\bigl[I_{\epsilon} +\gamma + \ln
  \epsilon\bigr] \stackrel{\gamma\to\infty}{=}
  -\frac{2}{(\alpha+\beta)^3}\,\ln \gamma \,.
\end{align}
The above equation yields the necessary condition for determining the missing
constant term in the general expression for the regularized integral $I_{\rm
  reg}$.

%
%
%
\begin{table*}[htb]
\caption{Nonrelativistic energies of helium-like ions $\cE^{(2)}_{\infty}$, $\cE^{(2)}_{M}$,
and $\cE^{(2)}_{M^2}$.  For the helium atom, the nonrelativistic energy
$\cE^{(2)}_{\infty}$  is given in the text, see Eqs.~(\ref{E2Hea})-(\ref{E2Heb}).
For $\cE^{(2)}_{\infty}$ and $\cE^{(2)}_{M}$, the results of fitting the numerical
data to the form $\cE = \sum_i c_i/Z^i$ are
also presented. The $1/Z$ expansion of the second-order recoil correction 
$\cE^{(2)}_{M^2}$ was not studied since this correction is relevant for the light 
atoms only.  Atomic units are used.
\label{tab:E2}}
\begin{ruledtabular}
  \begin{tabular}{l.....l}
    $Z$  &  \multicolumn{1}{c}{$1^1S$}
    &  \multicolumn{1}{c}{$2^1S$}
    &  \multicolumn{1}{c}{$2^3S$}
    &  \multicolumn{1}{c}{$2^1P$}
    &  \multicolumn{1}{c}{$2^3P$} \\
    \hline\\[-5pt]
    \multicolumn{5}{l}{$\cE^{(2)}_{\infty} + Z^2 (1+1/n^2)/2$}\\[2pt]
    3& 1.720\,086\,5x87\,330\,694 & 0.584\,123\,2x54\,404\,560&  0.514\,272\,6x27\,429\,260& 0.631\,648\,9x22\,219\,983&  0.597\,284\,3x18\,602\,632\\
    4& 2.344\,433\,7x61\,576\,413 & 0.815\,126\,1x04\,651\,679&  0.702\,833\,4x10\,222\,385& 0.889\,228\,3x77\,083\,556&  0.825\,026\,8x56\,929\,027\\
    5& 2.969\,028\,4x19\,757\,218 & 1.046\,471\,9x67\,118\,562&  0.891\,102\,6x51\,185\,767& 1.147\,716\,7x34\,692\,201&  1.051\,862\,3x07\,786\,520\\
    6& 3.593\,753\,3x98\,101\,470 & 1.277\,982\,2x98\,711\,029&  1.079\,244\,0x97\,692\,043& 1.406\,667\,6x86\,611\,591&  1.278\,289\,3x03\,511\,949\\
    7& 4.218\,554\,8x51\,227\,295 & 1.509\,584\,2x84\,500\,004&  1.267\,318\,2x62\,508\,963& 1.665\,883\,5x99\,383\,315&  1.504\,498\,2x55\,070\,109\\
    8& 4.843\,404\,8x77\,242\,074 & 1.741\,242\,6x92\,371\,423&  1.455\,352\,6x79\,914\,990& 1.925\,264\,7x64\,124\,369&  1.730\,577\,2x83\,616\,668\\
    9& 5.468\,287\,6x36\,040\,509 & 1.972\,938\,3x60\,878\,370&  1.643\,361\,6x70\,481\,692& 2.184\,755\,7x23\,292\,205&  1.956\,572\,7x06\,521\,290\\
    10& 6.093\,193\,4x84\,962\,451 & 2.204\,659\,9x53\,768\,328&  1.831\,353\,4x15\,926\,627& 2.444\,323\,2x60\,421\,919&  2.182\,511\,1x79\,215\,097\\
    11& 6.718\,116\,2x23\,927\,278 & 2.436\,400\,3x25\,538\,931&  2.019\,332\,9x29\,131\,276& 2.703\,946\,2x97\,095\,311&  2.408\,409\,1x21\,058\,957\\
    12& 7.343\,051\,6x87\,353\,070 & 2.668\,154\,7x43\,908\,730&  2.207\,303\,4x52\,397\,112& 2.963\,610\,8x24\,902\,483&  2.634\,277\,1x96\,002\,167  
    \\[5pt]
    \multicolumn{5}{l}{$1/Z$ expansion coefficients}\\[2pt]
    $c_{-1}$&      5/8x          &   169/x729        &   137/x729        &   1705x/6561      &   1481x/6561\\
    $c_{ 0}$&   -0.157\,x666\,43 &  -0.114\,x510\,14 &  -0.047\,x409\,30 &  -0.157\,x028\,66 &  -0.072\,x998\,98\\
    $c_{ 1}$&    0.008\,x699\,03 &   0.009\,x327\,61 &  -0.004\,x872\,28 &   0.026\,x106\,26 &  -0.016\,x585\,30\\
    $c_{ 2}$&   -0.000\,x888\,69 &  -0.001\,x284\,99 &  -0.003\,x457\,75 &   0.005\,x782\,46 &  -0.010\,x353\,67\\
    $c_{ 3}$&   -0.001\,x036\,59 &   0.006\,x194\,73 &  -0.002\,x030\,70 &  -0.005\,x033\,12 &  -0.005\,x427\,43\\
    $c_{ 4}$&   -0.000\,x610\,67 &  -0.001\,x471\,94 &  -0.001\,x278\,08 &  -0.007\,x099\,02 &  -0.002\,x001\,75\\
    $c_{ 5}$&   -0.000\,x388\,13 &  -0.003\,x775\,51 &  -0.000\,x934\,77 &  -0.001\,x103\,32 &   0.000\,x100\,74
    \\[5pt]
    \multicolumn{5}{l}{$\cE^{(2)}_M/(Z^2\,m/M)$}\\[2pt]
    2 &  0.765\,x698\,46 &  0.538\,x869\,48 &  0.545\,x667\,88 &  0.542\,x471\,90 &  0.517\,x147\,94\\
    3 &  0.840\,x987\,69 &  0.562\,x509\,01 &  0.569\,x810\,61 &  0.582\,x725\,90 &  0.524\,x732\,09\\
    4 &  0.879\,x755\,41 &  0.576\,x178\,76 &  0.582\,x830\,44 &  0.608\,x268\,18 &  0.529\,x299\,61\\
    5 &  0.903\,x348\,97 &  0.584\,x990\,29 &  0.590\,x909\,14 &  0.625\,x203\,60 &  0.532\,x349\,14\\
    6 &  0.919\,x210\,60 &  0.591\,x123\,83 &  0.596\,x399\,30 &  0.637\,x117\,06 &  0.534\,x516\,27\\
    7 &  0.930\,x603\,34 &  0.595\,x633\,60 &  0.600\,x370\,03 &  0.645\,x913\,88 &  0.536\,x130\,48\\
    8 &  0.939\,x181\,63 &  0.599\,x086\,96 &  0.603\,x374\,29 &  0.652\,x661\,02 &  0.537\,x377\,39\\
    9 &  0.945\,x873\,29 &  0.601\,x815\,29 &  0.605\,x726\,19 &  0.657\,x993\,92 &  0.538\,x368\,63\\
    10 &  0.951\,x238\,85 &  0.604\,x024\,79 &  0.607\,x617\,17 &  0.662\,x312\,12 &  0.539\,x175\,09\\
    11 &  0.955\,x636\,86 &  0.605\,x850\,39 &  0.609\,x170\,53 &  0.665\,x878\,55 &  0.539\,x843\,79\\
    12 &  0.959\,x307\,33 &  0.607\,x384\,02 &  0.610\,x469\,21 &  0.668\,x872\,98 &  0.540\,x407\,11
    \\[5pt]
    \multicolumn{5}{l}{$1/Z$ expansion coefficients}\\[2pt]
    $c_{0}$ &      1x          &       5/8x        &       5/8x       & 5/8+x2^9/3^8   & 5/8-x2^9/3^8\\
    $c_{1}$ & -0.491\,x706\,5  &  -0.219\,x681\,2  &  -0.176\,x992\,4 &-0.422\,x232\,8 &-0.082\,x563\,1\\
    $c_{2}$ &  0.039\,x651\,7  &   0.100\,x337\,6  &   0.030\,x684\,4 & 0.133\,x727\,0 & 0.045\,x923\,6\\
    $c_{3}$ &  0.012\,x972\,5  &  -0.009\,x874\,4  &   0.009\,x154\,5 & 0.162\,x310\,7 & 0.009\,x983\,7\\
    $c_{4}$ & -0.000\,x022\,6  &  -0.006\,x744\,6  &   0.003\,x896\,3 &-0.000\,x445\,3 &-0.007\,x945\,4\\
    $c_{5}$ &  0.003\,x037\,4  &   0.008\,x949\,1  &   0.001\,x835\,8 &-0.099\,x048\,5 &-0.011\,x326\,2
    \\[5pt]
    \multicolumn{5}{l}{$\cE^{(2)}_{M^2}/(Z^2\,m^2/M^2)$}\\[2pt]
    2 & -0.923\,x067\,75 &   -0.575\,x065\,28 &   -0.561\,x902\,54 &   -0.596\,x056\,62 &   -0.552\,x238\,02\\
    3 & -1.015\,x027\,87 &   -0.623\,x263\,22 &   -0.591\,x803\,40 &   -0.687\,x746\,68 &   -0.574\,x156\,42\\
    4 & -1.061\,x762\,41 &   -0.651\,x489\,97 &   -0.607\,x626\,64 &   -0.742\,x385\,92 &   -0.583\,x335\,75\\
    5 & -1.090\,x051\,34 &   -0.669\,x682\,17 &   -0.617\,x364\,45 &   -0.776\,x892\,85 &   -0.588\,x425\,34\\
    6 & -1.109\,x014\,61 &   -0.682\,x312\,57 &   -0.623\,x951\,64 &   -0.800\,x354\,77 &   -0.591\,x687\,77\\
    7 & -1.122\,x610\,79 &   -0.691\,x572\,88 &   -0.628\,x701\,72 &   -0.817\,x261\,02 &   -0.593\,x967\,08\\
    8 & -1.132\,x835\,81 &   -0.698\,x645\,59 &   -0.632\,x288\,26 &   -0.829\,x994\,71 &   -0.595\,x653\,31\\
    9 & -1.140\,x805\,12 &   -0.704\,x220\,80 &   -0.635\,x091\,75 &   -0.839\,x919\,81 &   -0.596\,x952\,97\\
    10 & -1.147\,x190\,96 &   -0.708\,x727\,06 &   -0.637\,x343\,21 &   -0.847\,x868\,27 &   -0.597\,x986\,12\\
    11 & -1.152\,x422\,61 &   -0.712\,x444\,12 &   -0.639\,x190\,98 &   -0.854\,x374\,66 &   -0.598\,x827\,53\\
    12 & -1.156\,x787\,03 &   -0.715\,x562\,17 &   -0.640\,x734\,67 &   -0.859\,x797\,49 &   -0.599\,x526\,27
  \end{tabular}
\end{ruledtabular}
\end{table*}

%
%
%
\begin{table*}[htb]
\caption{The leading relativistic corrections $\cE^{(4)}_{\infty}$ and
$\cE^{(4)}_{M}$ for helium-like atoms and their $1/Z$-expansion coefficients.  
The analytical results for the coefficient $c_1$ for $\cE^{(4)}_{\infty}$ 
were taken from Ref.~\cite{mohr:85:pra} for the $1^1S$,
$2^3S$, $2^3P_0$, and $2^3P_2$ states. For the other states, this
coefficient was evaluated numerically to high accuracy in this work by the same
method as in Ref.~\cite{mohr:85:pra}. The $c_0$ coefficient of
$\cE^{(4)}_{M}$ for the $S$ states originates from the one-electron recoil effect
and is well known from the hydrogen theory. For the $P$ states, it contains
also the two-electron contribution, which was derived in
Ref.~\cite{shabaev:94:rec}. The remaining $1/Z$-expansion coefficients
were obtained by fitting the numerical data for $\cE^{(4)}_{\infty}$ and
$\cE^{(4)}_{M}$. Atomic units are used.
\label{tab:E4}}
\begin{ruledtabular}
  \begin{tabular}{l.......}
    $Z$  &  \multicolumn{1}{c}{$1^1S$}
    &  \multicolumn{1}{c}{$2^1S$}
    &  \multicolumn{1}{c}{$2^3S$}
    &  \multicolumn{1}{c}{$2^1P_1$}
    &  \multicolumn{1}{c}{$2^3P_0$} 
    &  \multicolumn{1}{c}{$2^3P_1$} 
    &  \multicolumn{1}{c}{$2^3P_2$} \\
    \hline\\[-5pt]
    \multicolumn{5}{l}{$\cE^{(4)}_{\infty}/Z^4$}\\[2pt]
    2 &-0.121\,x984\,67 &-0.127\,x135\,46 &-0.135\,x279\,87 &-0.127\,x501\,60 &-0.118\,x042\,52 &-0.123\,x316\,23 &-0.123\,x729\,58 \\
    3 &-0.145\,x794\,73 &-0.131\,x881\,23 &-0.142\,x840\,30 &-0.130\,x953\,91 &-0.121\,x128\,29 &-0.126\,x601\,86 &-0.124\,x410\,72 \\
    4 &-0.163\,x263\,53 &-0.136\,x297\,75 &-0.147\,x356\,68 &-0.133\,x229\,53 &-0.126\,x667\,74 &-0.130\,x512\,05 &-0.125\,x569\,23 \\
    5 &-0.176\,x048\,64 &-0.139\,x882\,98 &-0.150\,x310\,08 &-0.134\,x786\,60 &-0.131\,x522\,71 &-0.133\,x708\,51 &-0.126\,x557\,82 \\
    6 &-0.185\,x674\,19 &-0.142\,x737\,56 &-0.152\,x383\,24 &-0.135\,x917\,90 &-0.135\,x444\,97 &-0.136\,x216\,54 &-0.127\,x342\,75 \\
    7 &-0.193\,x138\,21 &-0.145\,x028\,97 &-0.153\,x916\,10 &-0.136\,x778\,46 &-0.138\,x596\,36 &-0.138\,x199\,25 &-0.127\,x965\,82 \\
    8 &-0.199\,x077\,87 &-0.146\,x895\,44 &-0.155\,x094\,63 &-0.137\,x455\,97 &-0.141\,x156\,38 &-0.139\,x793\,29 &-0.128\,x467\,48 \\
    9 &-0.203\,x909\,05 &-0.148\,x439\,23 &-0.156\,x028\,56 &-0.138\,x003\,71 &-0.143\,x266\,17 &-0.141\,x097\,57 &-0.128\,x878\,10 \\
    10 &-0.207\,x911\,70 &-0.149\,x734\,47 &-0.156\,x786\,71 &-0.138\,x455\,96 &-0.145\,x029\,86 &-0.142\,x182\,11 &-0.129\,x219\,51 \\
    11 &-0.211\,x280\,06 &-0.150\,x835\,16 &-0.157\,x414\,35 &-0.138\,x835\,82 &-0.146\,x523\,58 &-0.143\,x096\,91 &-0.129\,x507\,41 \\
    12 &-0.214\,x152\,65 &-0.151\,x781\,23 &-0.157\,x942\,44 &-0.139\,x159\,46 &-0.147\,x803\,54 &-0.143\,x878\,26 &-0.129\,x753\,21 
    \\[5pt]
    \multicolumn{5}{l}{$1/Z$ expansion coefficients}\\[2pt]
    $c_{0}$ &   -1/x4         &   -21/x128       &   -21/x128       &   -55/x384         &   -21/x128   &   -59/x384   &   -17/x128   \\
    $c_{1}$ & 0.480\,x139\,61& 0.169\,x478\,18& 0.076\,x935\,23& 0.055\,x403\,03& 0.219\,x768\,22& 0.130\,x428\,76& 0.040\,x638\,72\\
    $c_{2}$ &-0.636\,x506\,86&-0.281\,x858\,62&-0.042\,x775\,47&-0.090\,x632\,15&-0.303\,x523\,35&-0.162\,x129\,41&-0.047\,x315\,68\\
    $c_{3}$ & 0.456\,x314\,23& 0.202\,x919\,21& 0.010\,x473\,95& 0.156\,x412\,39& 0.091\,x746\,25& 0.042\,x468\,90& 0.002\,x244\,38\\
    $c_{4}$ &-0.171\,x179\,61&-0.042\,x542\,10&-0.004\,x460\,83&-0.178\,x042\,53&-0.008\,x844\,33&-0.004\,x319\,44&-0.000\,x236\,51\\
    $c_{5}$ & 0.018\,x587\,49& 0.018\,x861\,71&-0.001\,x566\,73& 0.059\,x068\,31& 0.015\,x552\,82& 0.007\,x698\,46& 0.003\,x691\,05
    \\[5pt]
    \multicolumn{5}{l}{$\cE^{(4)}_{M}/(Z^4\,m/M)$}\\[2pt]
    2 & -0.134\,x960\,7 & -0.004\,x351\,6 &  0.005\,x574\,1 & -0.003\,x655\,3 &  0.015\,x596\,8 &  0.016\,x677\,1 &  0.012\,x760\,7\\
    3 & -0.123\,x759\,2 & -0.001\,x616\,1 &  0.011\,x426\,9 & -0.008\,x574\,4 &  0.026\,x148\,2 &  0.026\,x855\,2 &  0.019\,x248\,5\\
    4 & -0.107\,x627\,1 &  0.002\,x303\,9 &  0.015\,x288\,3 & -0.012\,x139\,6 &  0.032\,x666\,5 &  0.032\,x185\,5 &  0.021\,x650\,8\\
    5 & -0.093\,x784\,1 &  0.005\,x792\,4 &  0.017\,x933\,4 & -0.014\,x451\,6 &  0.037\,x580\,2 &  0.035\,x768\,1 &  0.022\,x926\,5\\
    6 & -0.082\,x625\,7 &  0.008\,x672\,2 &  0.019\,x840\,8 & -0.015\,x970\,4 &  0.041\,x407\,7 &  0.038\,x388\,7 &  0.023\,x736\,9\\
    7 & -0.073\,x647\,0 &  0.011\,x026\,9 &  0.021\,x276\,3 & -0.017\,x005\,1 &  0.044\,x450\,5 &  0.040\,x395\,3 &  0.024\,x304\,7\\
    8 & -0.066\,x337\,0 &  0.012\,x965\,9 &  0.022\,x393\,8 & -0.017\,x736\,6 &  0.046\,x915\,2 &  0.041\,x981\,4 &  0.024\,x727\,6\\
    9 & -0.060\,x299\,1 &  0.014\,x581\,1 &  0.023\,x287\,6 & -0.018\,x271\,0 &  0.048\,x946\,2 &  0.043\,x266\,3 &  0.025\,x056\,1\\
    10 & -0.055\,x241\,6 &  0.015\,x943\,0 &  0.024\,x018\,4 & -0.018\,x672\,8 &  0.050\,x645\,5 &  0.044\,x328\,0 &  0.025\,x319\,3\\
    11 & -0.050\,x950\,5 &  0.017\,x104\,5 &  0.024\,x626\,8 & -0.018\,x982\,2 &  0.052\,x086\,4 &  0.045\,x219\,7 &  0.025\,x535\,1\\
    12 & -0.047\,x267\,8 &  0.018\,x105\,5 &  0.025\,x141\,1 & -0.019\,x225\,6 &  0.053\,x322\,8 &  0.045\,x979\,1 &  0.025\,x715\,5
    \\[5pt]
    \multicolumn{5}{l}{$1/Z$ expansion coefficients}\\[2pt]
    $c_{0}$&       0\,x       &   1/x32         &    1/x32        & -0.020\,x744\,7 &  0.069\,x205\,9&   0.055\,x392\,0 &  0.027\,x764\,2\\
    $c_{1}$&  -0.645\,x040\,2 & -0.182\,x643\,4 & -0.078\,x412\,4 &  0.002\,x583\,0 & -0.217\,x113\,6&  -0.125\,x397\,2 & -0.025\,x605\,0\\
    $c_{2}$&   0.972\,x372\,8 &  0.314\,x800\,3 &  0.062\,x834\,3 &  0.220\,x104\,6 &  0.332\,x302\,1&   0.157\,x607\,0 &  0.015\,x453\,2\\
    $c_{3}$&  -0.460\,x091\,9 & -0.188\,x443\,6 & -0.018\,x866\,9 & -0.387\,x495\,1 & -0.152\,x038\,8&  -0.096\,x226\,5 & -0.036\,x663\,9\\
    $c_{4}$&  -0.040\,x368\,0 & -0.048\,x228\,2 &  0.000\,x413\,8 & -0.046\,x282\,4 & -0.252\,x878\,4&  -0.030\,x160\,4 & -0.017\,x642\,5
  \end{tabular}
\end{ruledtabular}
\end{table*}

%
%
%
\begin{table*}[htb]
\caption{
Bethe logarithm for helium-like atoms with the infinite nuclear mass,
$\ln(k_0/Z^2)$, and its first-order perturbation by the mass polarization
operator, $\ln(k_0)_M$. 
Coefficients of the $1/Z$ expansion of $\ln(k_0/Z^2)$ are also presented. 
The leading term $c_0$ is known with a high accuracy from the hydrogen
theory. The higher-order
coefficients are obtained by fitting the numerical data. 
\label{tab:bethelog}}
\begin{ruledtabular}
  \begin{tabular}{r.....l}
    $Z$  &  \multicolumn{1}{c}{$1^1S$}
    &  \multicolumn{1}{c}{$2^1S$}
    &  \multicolumn{1}{c}{$2^3S$}
    &  \multicolumn{1}{c}{$2^1P$}
    &  \multicolumn{1}{c}{$2^3P$} 
    &  \multicolumn{1}{c}{Ref.}\\
    \hline\\[-5pt]
    \multicolumn{5}{l}{$\ln(k_0/Z^2)$}\\[2pt]
    2 & 2.983\,8x65\,861\,8\,(1)& 2.980\,11x8\,365\,1\,(1)& 2.977\,7x42\,459\,29\,(2)& 2.983\,8x03\,382\,4\,(1)& 2.983\,6x91\,003\,3\,(2)&  \\
    & 2.983\,8x65\,860\,9\,(1)& 2.980\,11x8\,364\,8\,(1)& 2.977\,7x42\,459\,2\,(1) & 2.983\,8x03\,377\,(1)   &  2.983\,6x90\,995\,(1)  &\cite{korobov:04} \\
    & 2.983\,8x65\,857\,(3)   & 2.980\,11x8\,36\,(7)    & 2.977\,7x42\,46\,(1)     & 2.983\,8x03\,46\,(3)    &  2.983\,6x90\,84\,(2)   &\cite{drake:99:cjp}\\
    3 & 2.982\,6x24\,563\,0\,(2)& 2.976\,36x3\,063\,0\,(2)& 2.973\,8x51\,709\,92\,(4)& 2.983\,1x86\,013\,6\,(2) & 2.982\,9x58\,798\,2\,(2)&  \\
    4 & 2.982\,5x03\,099\,1\,(3)& 2.973\,97x6\,911\,2\,(3)& 2.971\,7x35\,578\,90\,(7)& 2.982\,6x98\,213\,8\,(4) & 2.982\,4x43\,598\,4\,(3)&  \\
    5 & 2.982\,5x91\,376\,1\,(4)& 2.972\,38x8\,098\,8\,(4)& 2.970\,4x24\,964\,90\,(8)& 2.982\,3x40\,114\,9\,(8) & 2.982\,0x89\,604\,9\,(4)&  \\
    6 & 2.982\,7x16\,948\,(1)   & 2.971\,26x6\,246\,4\,(5)& 2.969\,5x37\,071\,9\,(3) & 2.982\,0x72\,719\,(2)   &  2.981\,8x35\,938\,5\,(6)&  \\
    & 2.982\,7x16\,948\,(4)   & 2.971\,26x6\,24\,(4)    & 2.969\,5x37\,07\,(1)    &  2.982\,0x72\,76\,(2)    &  2.981\,8x35\,92\,(3)   &\cite{drake:99:cjp}\\
    7 & 2.982\,8x39\,085\,(3)   & 2.970\,43x5\,367\,(1)   & 2.968\,8x96\,814\,(1)   &  2.981\,8x67\,337\,(7)   &  2.981\,6x46\,451\,(2)  &  \\
    8 & 2.982\,9x48\,318\,(4)   & 2.969\,79x6\,528\,(3)   & 2.968\,4x13\,645\,(2)   &  2.981\,7x05\,33\,(1)    &  2.981\,4x99\,939\,(4)  &  \\
    9 & 2.983\,0x43\,667\,(8)   & 2.969\,29x0\,586\,(5)   & 2.968\,0x36\,227\,(5)   &  2.981\,5x74\,56\,(3)    &  2.981\,3x83\,443\,(7)  &  \\
    10 & 2.983\,1x26\,46\,(2)    & 2.968\,88x0\,24\,(1)    & 2.967\,7x33\,341\,(9)   &  2.981\,4x66\,92\,(5)    &  2.981\,2x88\,68\,(1)   &  \\
    11 & 2.983\,1x98\,50\,(3)    & 2.968\,54x0\,85\,(2)    & 2.967\,4x84\,93\,(2)    &  2.981\,3x76\,9\,(1)     &  2.981\,2x10\,12\,(2)   &  \\
    12 & 2.983\,2x61\,47\,(5)    & 2.968\,25x5\,57\,(4)    & 2.967\,2x77\,54\,(3)    &  2.981\,3x00\,4\,(2)     &  2.981\,1x43\,96\,(3)   &  
    \\[5pt]
    \multicolumn{5}{l}{Coefficients of the $1/Z$ expansion:}\\[2pt]
    $c_0$ &   2.984\,1x28\,56  &  2.964\,9x77\,59 &   2.964\,9x77\,59 &   2.980\,3x76\,47  &  2.980\,3x76\,47 \\
    $c_1$ &  -0.012\,2x99\,28  &  0.040\,7x88\,09 &   0.027\,7x59\,43 &   0.012\,0x03\,83  &  0.009\,6x27\,97 \\
    $c_2$ &   0.022\,4x49\,74  & -0.016\,4x39\,35 &  -0.001\,4x23\,95 &  -0.010\,9x82\,08  & -0.004\,8x10\,60 \\
    $c_3$ &   0.003\,5x86\,19  & -0.012\,3x55\,03 &  -0.005\,9x68\,56 &  -0.000\,4x82\,19  & -0.002\,4x57\,88 \\
    $c_4$ &  -0.002\,5x03\,70  &  0.005\,8x13\,30 &   0.000\,1x19\,95 &   0.003\,7x64\,32  & -0.000\,2x36\,70 
    \\[5pt]
    \multicolumn{5}{l}{$\ln(k_0)_M/(m/M)$}\\[2pt]
    2  &   0.094\,3x89\,4\,(1)  &  0.017\,7x34\,4\,(1) &  0.004\,7x85\,54\,(1) & -0.003\,x553\,4\,(2)& 0.008\,7x09\,5\,(1) &\\
    &   0.094\,3x8\,(1)      &  0.017\,7x34\,(1)    &  0.004\,7x84\,(3)     & -0.003\,x538\,(6)   & 0.008\,7x01\,(4)
    &\cite{drake:99:cjp}\\
    3  &   0.109\,5x39\,7\,(1)  &  0.034\,2x10\,3\,(1) &  0.007\,8x52\,51\,(1) & -0.006\,x602\,3\,(2)& 0.016\,3x28\,3\,(1) &\\     
    4  &   0.116\,9x19\,7\,(1)  &  0.044\,8x76\,8\,(1) &  0.009\,6x16\,61\,(1) & -0.007\,x951\,2\,(2)& 0.020\,1x99\,2\,(1) &\\     
    5  &   0.121\,3x04\,5\,(2)  &  0.052\,0x12\,4\,(2) &  0.010\,7x54\,20\,(1) & -0.008\,x629\,5\,(2)& 0.022\,4x79\,0\,(1) &\\     
    6  &   0.124\,2x12\,9\,(3)  &  0.057\,0x53\,0\,(3) &  0.011\,5x47\,92\,(1) & -0.009\,x015\,3\,(2)& 0.023\,9x73\,6\,(1) &\\      
    &   0.124\,2x1\,(1)      &  0.057\,0x51\,(1)    &  0.011\,5x41\,(1)     & -0.008\,x98\,(1)    & 0.023\,9x8\,(1)
    &\cite{drake:99:cjp}\\
    7  &   0.126\,2x83\,1\,(4)  &  0.060\,7x83\,0\,(9) &  0.012\,1x33\,20\,(1) & -0.009\,x255\,6\,(2)& 0.025\,0x27\,3\,(1) &\\      
    8  &   0.127\,8x33\,6\,(5)  &  0.063\,6x47\,0\,(7) &  0.012\,5x82\,68\,(1) & -0.009\,x416\,0\,(2)& 0.025\,8x09\,5\,(2) &\\      
    9  &   0.129\,0x37\,1\,(2)  &  0.065\,9x12\,(1)    &  0.012\,9x38\,76\,(1) & -0.009\,x528\,7\,(2)& 0.026\,4x12\,8\,(3) &\\      
    10  &   0.129\,9x98\,9\,(2)  &  0.067\,7x46\,(1)    &  0.013\,2x27\,87\,(1) & -0.009\,x611\,2\,(2)& 0.026\,8x92\,2\,(4) &\\      
    11  &   0.130\,7x85\,1\,(2)  &  0.069\,2x61\,(1)    &  0.013\,4x67\,30\,(1) & -0.009\,x673\,6\,(2)& 0.027\,2x82\,2\,(4) &\\      
    12  &   0.131\,4x39\,7\,(2)  &  0.070\,5x34\,(2)    &  0.013\,6x68\,86\,(1) & -0.009\,x722\,2\,(3)& 0.027\,6x05\,7\,(4) &      
  \end{tabular}
\end{ruledtabular}
\end{table*}

%
%
%
\begin{table*}[htb]
\caption{The leading logarithmic QED corrections $\cE^{(5)}_{\infty}({\rm log})$
and $\cE^{(5)}_{M}({\rm log})$.
For the non-recoil correction, we present the coefficients of the $1/Z$
expansion obtained by fitting the numerical data (except for $c_0$ which
is known analytically). The recoil
correction is very small for ions with $Z>12$, so its $1/Z$ expansion was not studied.
Atomic units are used.
\label{tab:E5log}}
\begin{ruledtabular}
  \begin{tabular}{l.....l}
    $Z$  &  \multicolumn{1}{c}{$1^1S$}
    &  \multicolumn{1}{c}{$2^1S$}
    &  \multicolumn{1}{c}{$2^3S$}
    &  \multicolumn{1}{c}{$2^1P$}
    &  \multicolumn{1}{c}{$2^3P$} \\
    \hline\\[-5pt]
    \multicolumn{5}{l}{$\cE^{(5)}_{\infty}({\rm log})/[Z^4\,\ln(\Za)^{-2}]$}\\[2pt]
    2 &  0.587\,x967\,740 &  0.435\,x225\,697 &  0.440\,x118\,361 &  0.424\,x690\,417 &  0.419\,x620\,202 \\
    3 &  0.661\,x366\,949 &  0.444\,x453\,845 &  0.450\,x745\,298 &  0.425\,x056\,974 &  0.418\,x277\,856 \\ 
    4 &  0.702\,x709\,966 &  0.450\,x600\,513 &  0.456\,x788\,812 &  0.425\,x087\,644 &  0.418\,x654\,616 \\ 
    5 &  0.729\,x153\,414 &  0.454\,x883\,655 &  0.460\,x628\,756 &  0.425\,x031\,723 &  0.419\,x253\,068 \\ 
    6 &  0.747\,x506\,286 &  0.458\,x013\,834 &  0.463\,x274\,004 &  0.424\,x963\,548 &  0.419\,x808\,283 \\ 
    7 &  0.760\,x984\,320 &  0.460\,x392\,952 &  0.465\,x204\,144 &  0.424\,x901\,229 &  0.420\,x280\,937 \\ 
    8 &  0.771\,x300\,325 &  0.462\,x259\,020 &  0.466\,x673\,600 &  0.424\,x848\,038 &  0.420\,x676\,712 \\ 
    9 &  0.779\,x449\,465 &  0.463\,x760\,328 &  0.467\,x829\,292 &  0.424\,x803\,420 &  0.421\,x008\,829 \\ 
    10 &  0.786\,x049\,205 &  0.464\,x993\,553 &  0.468\,x761\,819 &  0.424\,x766\,013 &  0.421\,x289\,730 \\ 
    11 &  0.791\,x502\,875 &  0.466\,x024\,214 &  0.469\,x530\,027 &  0.424\,x734\,472 &  0.421\,x529\,561 \\ 
    12 &  0.796\,x085\,045 &  0.466\,x898\,202 &  0.470\,x173\,774 &  0.424\,x707\,664 &  0.421\,x736\,263  
    \\[5pt]
    \multicolumn{5}{l}{$1/Z$ expansion coefficients}\\[2pt]
    $c_{0}$&    8/(3x\pi)   &   3/(2x\pi)   &   3/(2x\pi)   &   4/(3x\pi)   &   4/(3x\pi)\\
    $c_{1}$&   -0.659\,x550\,48 &  -0.137\,x744\,61 &  -0.089\,x756\,44 &   0.003\,x158\,46 &  -0.036\,x478\,76\\
    $c_{2}$&    0.330\,x586\,16 &   0.136\,x452\,03 &   0.026\,x693\,63 &   0.009\,x117\,22 &   0.051\,x362\,86\\
    $c_{3}$&   -0.132\,x768\,16 &  -0.061\,x297\,50 &   0.005\,x488\,16 &  -0.061\,x231\,69 &   0.011\,x132\,17\\
    $c_{4}$&    0.048\,x550\,42 &  -0.000\,x485\,58 &   0.001\,x979\,36 &   0.070\,x367\,56 &  -0.001\,x924\,41\\
    $c_{5}$&   -0.005\,x834\,76 &  -0.000\,x419\,68 &   0.001\,x082\,09 &   0.002\,x994\,02 &  -0.014\,x225\,78\\
    \\[5pt]
    \multicolumn{5}{l}{$\cE^{(5)}_{M}({\rm log})/[(m/M)\,Z^4\,\ln(\Za)^{-2}]$}\\[2pt]
    2 & -1.490\,x787\,8 & -1.087\,x827\,3 & -1.099\,x571\,6 & -1.048\,x033\,4 & -1.072\,x942\,1 \\
    3 & -1.503\,x000\,7 & -0.999\,x820\,0 & -1.013\,x525\,6 & -0.931\,x087\,7 & -0.972\,x929\,1 \\
    4 & -1.414\,x815\,1 & -0.900\,x768\,2 & -0.913\,x023\,5 & -0.820\,x369\,5 & -0.869\,x549\,4 \\
    5 & -1.281\,x038\,5 & -0.795\,x295\,8 & -0.805\,x624\,3 & -0.712\,x358\,9 & -0.765\,x222\,3 \\
    6 & -1.122\,x852\,5 & -0.685\,x953\,6 & -0.694\,x495\,7 & -0.605\,x509\,9 & -0.660\,x414\,8 \\
    7 & -0.950\,x101\,1 & -0.574\,x135\,7 & -0.581\,x137\,5 & -0.499\,x185\,8 & -0.555\,x305\,8 \\
    8 & -0.767\,x968\,8 & -0.460\,x648\,8 & -0.466\,x344\,1 & -0.393\,x108\,1 & -0.449\,x990\,7 \\
    9 & -0.579\,x443\,0 & -0.345\,x987\,0 & -0.350\,x573\,6 & -0.287\,x146\,9 & -0.344\,x527\,5 \\
    10 & -0.386\,x365\,2 & -0.230\,x468\,4 & -0.234\,x108\,4 & -0.181\,x238\,4 & -0.238\,x954\,2 \\
    11 & -0.189\,x932\,1 & -0.114\,x306\,4 & -0.117\,x131\,8 & -0.075\,x350\,3 & -0.133\,x296\,7 \\
    12 &  0.009\,x045\,2 &  0.002\,x350\,7 &  0.000\,x232\,1 &  0.030\,x534\,1 & -0.027\,x573\,6 \\
  \end{tabular}
\end{ruledtabular}
\end{table*}

%
%
%
\begin{table*}[htb]
\caption{The leading nonlogarithmic QED corrections $\cE^{(5)}_{\infty}({\rm nlog})$
and $\cE^{(5)}_{M}({\rm nlog})$.
For the non-recoil part, we present the coefficients of the $1/Z$
expansion. The coefficient $c_0$ is known with a
very good accuracy from the hydrogen theory. The remaining coefficients were
obtained by numerical fitting. The radiative recoil
correction is very small for ions with $Z>12$, so its $1/Z$ expansion was not
studied. Atomic units are used.
\label{tab:E5nlog}}
\begin{ruledtabular}
  \begin{tabular}{l.......}
    $Z$  &  \multicolumn{1}{c}{$1^1S$}
    &  \multicolumn{1}{c}{$2^1S$}
    &  \multicolumn{1}{c}{$2^3S$}
    &  \multicolumn{1}{c}{$2^1P_1$}
    &  \multicolumn{1}{c}{$2^3P_0$} 
    &  \multicolumn{1}{c}{$2^3P_1$} 
    &  \multicolumn{1}{c}{$2^3P_2$} \\
    \hline\\[-5pt]
    \multicolumn{5}{l}{$\cE^{(5)}_{\infty}({\rm nlog})/Z^4$}\\[2pt]
    2 & -1.390\,x282\,4 & -1.021\,x756\,7 & -1.032\,x719\,5 & -0.999\,x104\,6 & -0.986\,x487\,5 & -0.987\,x824\,8 & -0.987\,x273\,6 \\
    3 & -1.552\,x423\,4 & -1.040\,x733\,9 & -1.055\,x874\,3 & -1.000\,x028\,8 & -0.984\,x014\,5 & -0.985\,x285\,6 & -0.983\,x645\,7 \\
    4 & -1.645\,x829\,1 & -1.053\,x494\,3 & -1.068\,x951\,4 & -0.999\,x838\,9 & -0.985\,x557\,5 & -0.986\,x317\,8 & -0.983\,x817\,7 \\
    5 & -1.706\,x668\,4 & -1.062\,x551\,3 & -1.077\,x230\,3 & -0.999\,x427\,5 & -0.987\,x480\,3 & -0.987\,x748\,5 & -0.984\,x617\,0 \\
    6 & -1.749\,x464\,5 & -1.069\,x274\,8 & -1.082\,x920\,6 & -0.999\,x028\,5 & -0.989\,x184\,7 & -0.989\,x040\,9 & -0.985\,x437\,6 \\
    7 & -1.781\,x211\,9 & -1.074\,x448\,4 & -1.087\,x066\,2 & -0.998\,x687\,5 & -0.990\,x609\,4 & -0.990\,x128\,4 & -0.986\,x162\,1 \\
    8 & -1.805\,x701\,0 & -1.078\,x546\,0 & -1.090\,x218\,7 & -0.998\,x404\,7 & -0.991\,x791\,0 & -0.991\,x033\,0 & -0.986\,x779\,7 \\
    9 & -1.825\,x165\,8 & -1.081\,x868\,4 & -1.092\,x695\,9 & -0.998\,x170\,7 & -0.992\,x776\,8 & -0.991\,x788\,8 & -0.987\,x303\,6 \\
    10 & -1.841\,x008\,7 & -1.084\,x614\,8 & -1.094\,x693\,4 & -0.997\,x975\,9 & -0.993\,x607\,6 & -0.992\,x426\,0 & -0.987\,x749\,6 \\
    11 & -1.854\,x154\,5 & -1.086\,x922\,2 & -1.096\,x338\,0 & -0.997\,x812\,2 & -0.994\,x315\,0 & -0.992\,x968\,8 & -0.988\,x132\,3 \\
    12 & -1.865\,x237\,8 & -1.088\,x887\,5 & -1.097\,x715\,6 & -0.997\,x673\,3 & -0.994\,x923\,5 & -0.993\,x435\,8 & -0.988\,x463\,2 
    \\[5pt]
    \multicolumn{5}{l}{$1/Z$ expansion coefficients}\\[2pt]
    $c_{0}$ & -1.995\,x417\,0 &-1.113\,x278\,1 &-1.113\,x278\,1 &-0.996\,x116\,0 &-1.002\,x747\,5 &-0.999\,x431\,8 &-0.992\,x800\,3\\
    $c_{1}$ &  1.658\,x816\,0 & 0.325\,x517\,0 & 0.191\,x147\,5 &-0.017\,x559\,8 & 0.106\,x210\,2 & 0.081\,x192\,9 & 0.059\,x452\,5\\
    $c_{2}$ & -1.226\,x271\,4 &-0.420\,x456\,0 &-0.051\,x603\,3 &-0.031\,x853\,8 &-0.146\,x537\,2 &-0.108\,x500\,3 &-0.085\,x893\,4\\
    $c_{3}$ &  0.825\,x843\,5 & 0.327\,x325\,0 &-0.013\,x194\,5 & 0.247\,x256\,7 &-0.016\,x628\,5 &-0.029\,x527\,6 &-0.036\,x773\,9\\
    $c_{4}$ & -0.373\,x062\,3 &-0.114\,x627\,2 &-0.008\,x949\,1 &-0.334\,x953\,5 & 0.006\,x846\,2 & 0.008\,x201\,3 & 0.009\,x431\,2\\
    $c_{5}$ &  0.088\,x160\,0 & 0.037\,x575\,2 & 0.010\,x984\,6 & 0.085\,x299\,0 & 0.018\,x205\,0 & 0.016\,x141\,2 & 0.015\,x502\,4
    \\[5pt]
    \multicolumn{5}{l}{$\cE^{(5)}_{M}({\rm nlog})/[(m/M)\, Z^5]$}\\[2pt]
    2 &  3.292\,x520 &  2.455\,x583 &  2.489\,x805 &  2.385\,x181 &  2.393\,x644 &  2.393\,x984 &  2.393\,x698 \\
    3 &  2.816\,x621 &  1.914\,x328 &  1.949\,x969 &  1.811\,x294 &  1.817\,x918 &  1.818\,x182 &  1.817\,x751 \\
    4 &  2.527\,x383 &  1.642\,x421 &  1.673\,x220 &  1.526\,x466 &  1.531\,x293 &  1.531\,x426 &  1.530\,x930 \\
    5 &  2.334\,x128 &  1.478\,x228 &  1.504\,x411 &  1.356\,x337 &  1.359\,x930 &  1.359\,x969 &  1.359\,x460 \\
    6 &  2.196\,x193 &  1.368\,x117 &  1.390\,x605 &  1.243\,x181 &  1.245\,x918 &  1.245\,x898 &  1.245\,x400 \\
    7 &  2.092\,x908 &  1.289\,x066 &  1.308\,x659 &  1.162\,x444 &  1.164\,x573 &  1.164\,x516 &  1.164\,x039 \\
    8 &  2.012\,x719 &  1.229\,x525 &  1.246\,x828 &  1.101\,x918 &  1.103\,x606 &  1.103\,x526 &  1.103\,x074 \\
    9 &  1.948\,x679 &  1.183\,x048 &  1.198\,x512 &  1.054\,x847 &  1.056\,x206 &  1.056\,x113 &  1.055\,x685 \\
    10 &  1.896\,x366 &  1.145\,x753 &  1.159\,x714 &  1.017\,x186 &  1.018\,x295 &  1.018\,x194 &  1.017\,x790 \\
    11 &  1.852\,x835 &  1.115\,x160 &  1.127\,x873 &  0.986\,x367 &  0.987\,x283 &  0.987\,x177 &  0.986\,x794 \\
    12 &  1.816\,x049 &  1.089\,x607 &  1.101\,x272 &  0.960\,x678 &  0.961\,x441 &  0.961\,x334 &  0.960\,x971 \\
  \end{tabular}
\end{ruledtabular}
\end{table*}

%
%
%
\begin{table*}[htb]
\caption{The $m\alpha^6$ corrections $\cE^{(6)}_{\infty}({\rm
    log})$ and $\cE^{(6)}_{\infty}({\rm nlog})$
and their $1/Z$ expansion coefficients. Atomic units are used.
\label{tab:E6}}
\begin{ruledtabular}
  \begin{tabular}{l.......}
    $Z$  &  \multicolumn{1}{c}{$1^1S$}
    &  \multicolumn{1}{c}{$2^1S$}
    &  \multicolumn{1}{c}{$2^3S$}
    &  \multicolumn{1}{c}{$2^1P_1$}
    &  \multicolumn{1}{c}{$2^3P_0$} 
    &  \multicolumn{1}{c}{$2^3P_1$} 
    &  \multicolumn{1}{c}{$2^3P_2$} \\
    \hline\\[-5pt]
    \multicolumn{3}{l}{$\cE^{(6)}_{\infty}({\rm log})/[Z^3\,\ln(\Za)^{-2}]$}\\[2pt]
    2 &  0.020x\,880\,865 &  0.001x\,698\,116 & 0x & 0.000x\,144\,350 & 0x& 0x& 0x\\
    3 &  0.031x\,050\,719 &  0.003x\,738\,928 & 0x & 0.000x\,572\,351 & 0x& 0x& 0x\\
    4 &  0.037x\,377\,475 &  0.005x\,250\,650 & 0x & 0.001x\,013\,004 & 0x& 0x& 0x\\
    5 &  0.041x\,625\,375 &  0.006x\,344\,104 & 0x & 0.001x\,386\,910 & 0x& 0x& 0x\\
    6 &  0.044x\,658\,932 &  0.007x\,157\,223 & 0x & 0.001x\,692\,317 & 0x& 0x& 0x\\
    7 &  0.046x\,928\,852 &  0.007x\,781\,230 & 0x & 0.001x\,941\,610 & 0x& 0x& 0x\\
    8 &  0.048x\,689\,367 &  0.008x\,273\,618 & 0x & 0.002x\,147\,092 & 0x& 0x& 0x\\
    9 &  0.050x\,093\,825 &  0.008x\,671\,365 & 0x & 0.002x\,318\,555 & 0x& 0x& 0x\\
    10 &  0.051x\,239\,918 &  0.008x\,999\,033 & 0x &  0.002x\,463\,398 & 0x& 0x& 0x\\
    11 &  0.052x\,192\,721 &  0.009x\,273\,471 & 0x &  0.002x\,587\,159 & 0x& 0x& 0x\\
    12 &  0.052x\,997\,208 &  0.009x\,506\,579 & 0x &  0.002x\,694\,007 & 0x& 0x& 0x
    \\[5pt]
    \multicolumn{3}{l}{$1/Z$ expansion coefficients}\\[2pt]
    $c_{0}$&    1/x16      &   1/x81      & 0x &  1/2x43      & 0x& 0x& 0x\\
    $c_{1}$&   -0.121x\,468\,0 &  -0.037x\,197\,7 & 0x & -0.019x\,997\,1 & 0x& 0x& 0x\\
    $c_{2}$&    0.091x\,963\,7 &   0.038x\,711\,4 & 0x &  0.037x\,995\,7 & 0x& 0x& 0x\\
    $c_{3}$&   -0.033x\,400\,9 &  -0.014x\,283\,2 & 0x & -0.032x\,995\,1 & 0x& 0x& 0x\\
    $c_{4}$&    0.004x\,804\,4 &   0.003x\,600\,0 & 0x &  0.008x\,865\,3 & 0x& 0x& 0x\\
    $c_{5}$&   -0.000x\,487\,4 &  -0.006x\,272\,6 & 0x &  0.002x\,867\,7 & 0x& 0x& 0x
    \\[5pt]
    \multicolumn{5}{l}{$\cE^{(6)}_{\infty}({\rm nlog})/Z^6$}\\[2pt]
    2& 2.181x\,233\,3\,(1)& 1.528x\,981\,(2)& 1.536x\,593\,1\,(1)& 1.489x\,195\,(1) & 1.459x\,456\,(1) & 1.460x\,802\,(1) & 1.466x\,251\,(1)\\  
    3& 1.582x\,471\,7\,(1)& 1.016x\,337\,(1)& 1.020x\,440\,5\,(1)& 0.977x\,073\,(1) & 0.945x\,154\,(1) & 0.937x\,827\,(1) & 0.950x\,303\,(1)\\   
    4& 1.224x\,451\,9\,(8)& 0.753x\,467\,(1)& 0.754x\,660\,8\,(1)& 0.723x\,937\,(1) & 0.691x\,625\,(1) & 0.676x\,327\,(1) & 0.696x\,280\,(1)\\   
    5& 0.989x\,185\,0\,(1)& 0.592x\,746\,(1)& 0.592x\,238\,5\,(1)& 0.574x\,526\,(1) & 0.539x\,381\,(1) & 0.517x\,140\,(1) & 0.544x\,513\,(1)\\   
    6& 0.823x\,348\,1\,(1)& 0.484x\,094\,(1)& 0.482x\,626\,2\,(4)& 0.477x\,044\,(1) & 0.437x\,493\,(1) & 0.408x\,810\,(1) & 0.443x\,440\,(1)\\  
    7& 0.700x\,317\,8\,(1)& 0.405x\,658\,(1)& 0.403x\,650\,3\,(1)& 0.409x\,248\,(1) & 0.364x\,415\,(1) & 0.329x\,514\,(1) & 0.371x\,247\,(1)\\  
    8& 0.605x\,469\,6\,(1)& 0.346x\,341\,(1)& 0.344x\,035\,2\,(1)& 0.360x\,008\,(1) & 0.309x\,399\,(1) & 0.268x\,374\,(1) & 0.317x\,082\,(1) \\  
    9& 0.530x\,138\,1\,(1)& 0.299x\,896\,(1)& 0.297x\,436\,0\,(1)& 0.323x\,136\,(1) & 0.266x\,468\,(1) & 0.219x\,353\,(1) & 0.274x\,935\,(1) \\  
    10& 0.468x\,871\,6\,(1)& 0.262x\,537\,(1)& 0.260x\,008\,6\,(1)& 0.294x\,919\,(1) & 0.232x\,024\,(1) & 0.178x\,826\,(1) & 0.241x\,202\,(1) \\  
    11& 0.418x\,072\,5\,(6)& 0.231x\,831\,(1)& 0.229x\,287\,1\,(1)& 0.272x\,996\,(1) & 0.203x\,772\,(2) & 0.144x\,482\,(4) & 0.213x\,589\,(1)\\
    12& 0.375x\,272\,1\,(1)& 0.206x\,144\,(1)& 0.203x\,617\,0\,(4)& 0.255x\,791\,(1) & 0.180x\,178\,(2) & 0.114x\,783\,(4) & 0.190x\,568\,(1)
    \\[5pt]
    \multicolumn{5}{l}{$1/Z$ expansion coefficients}\\[2pt]
    $c_{-1}$&     0\ x     &      0x     &     0x      &  729/1x14688&     0x      & -729/1x14688&     0x     \\
    $c_{ 0}$& -1/8  x & -85/1x024   & -85/1x024   & -0.079x2398 & -85/1x024   & -0.067x2446 & -65/1x024    \\
    $c_{ 1}$& 6.342x\,8979 &  3.549x\,6121 &  3.487x\,5483 &  3.099x\,80   &  3.204x\,5132 &  3.119x\,05   &  3.059x\,7402\\
    $c_{ 2}$&-4.261x\,9    & -1.042x\,6    & -0.590x\,0    &  0.068x\,6    & -0.623x\,6    & -0.260x\,9    & -0.157x\,8\\
    $c_{ 3}$& 2.424x\,1    &  1.087x\,0    &  0.155x\,7    & -0.039x\,0    &  0.833x\,4    &  0.328x\,9    &  0.271x\,4\\
    $c_{ 4}$&-2.408x\,8    & -0.815x\,7    &  0.053x\,9    & -0.401x\,3    & -0.210x\,5    &  0.071x\,7    &  0.101x\,0
  \end{tabular}
\end{ruledtabular}
\end{table*}

%
%
%
\begin{table*}[htb]
\caption{The finite nuclear size correction $E_{\rm fs}$
 (the used values of the root-mean-square nuclear charge
  radii are listed in Table~\ref{tab:total}) and the higher-order correction $\cE^{(7+)}
  \equiv \cE^{(7+)}_{\rm rad}+ \cE^{(7+)}_{\rm nrad}$. Contributions to the 
  {\em ionization energy} are presented. Numerical values of the finite
 nuclear size correction are scaled by the same factor as for the higher-order
 correction, in order to simplify the comparison between them.
  \label{tab:E7}}
\begin{ruledtabular}
  \begin{tabular}{l........}
    $Z$  
    &  \multicolumn{1}{c}{$1^1S$}
    &  \multicolumn{1}{c}{$2^1S$}
    &  \multicolumn{1}{c}{$2^3S$}
    &  \multicolumn{1}{c}{$2^1P_1$}
    &  \multicolumn{1}{c}{$2^3P_0$} 
    &  \multicolumn{1}{c}{$2^3P_1$} 
    &  \multicolumn{1}{c}{$2^3P_2$} \\
    \hline\\[-5pt]
    \multicolumn{3}{l}{$E_{\rm fs}/[m\alpha^7Z^6]$}\\[2pt]
 2 & 3.4x1\,(1)      & 0.23x46\,(5)    & 0.30x39\,(7)    & 0.00x735\,(2)   & -0.09x14\,(3)   & -0.09x14\,(3)   & -0.09x14\,(3)  \\
 3 & 4.5x\,(1)       & 0.39x3\,(8)     & 0.47x\,(1)      & 0.01x66\,(4)    & -0.11x0\,(3)    & -0.11x0\,(3)    & -0.11x0\,(3)   \\
 4 & 3.1x6\,(3)      & 0.30x5\,(3)     & 0.35x1\,(3)     & 0.01x14\,(1)    & -0.06x23\,(6)   & -0.06x22\,(6)   & -0.06x23\,(6)  \\
 5 & 2.0x1\,(5)      & 0.20x6\,(5)     & 0.23x0\,(6)     & 0.00x66\,(2)    & -0.03x27\,(8)   & -0.03x27\,(8)   & -0.03x27\,(8)  \\
 6 & 1.5x60\,(4)     & 0.16x55\,(5)    & 0.18x18\,(5)    & 0.00x455\,(1)   & -0.02x150\,(6)  & -0.02x144\,(6)  & -0.02x150\,(6) \\
 7 & 1.2x81\,(8)     & 0.13x96\,(8)    & 0.15x12\,(9)    & 0.00x336\,(2)   & -0.01x528\,(9)  & -0.01x521\,(9)  & -0.01x528\,(9) \\
 8 & 1.1x30\,(6)     & 0.12x56\,(7)    & 0.13x47\,(7)    & 0.00x268\,(1)   & -0.01x187\,(6)  & -0.01x180\,(6)  & -0.01x187\,(6) \\
 9 & 1.0x56\,(5)     & 0.11x91\,(6)    & 0.12x68\,(6)    & 0.00x229\,(1)   & -0.00x990\,(5)  & -0.00x981\,(5)  & -0.00x990\,(5) \\
10 & 0.9x42\,(5)     & 0.10x76\,(6)    & 0.11x38\,(6)    & 0.00x188\,(1)   & -0.00x797\,(4)  & -0.00x788\,(4)  & -0.00x797\,(4) \\
11 & 0.7x89\,(5)     & 0.09x11\,(6)    & 0.09x58\,(6)    & 0.00x146\,(1)   & -0.00x608\,(4)  & -0.00x599\,(4)  & -0.00x608\,(4) \\
12 & 0.7x05\,(5)     & 0.08x21\,(6)    & 0.08x60\,(7)    & 0.00x1219\,(9)  & -0.00x499\,(4)  & -0.00x490\,(4)  & -0.00x499\,(4) 
    \\[5pt]
    \multicolumn{5}{l}{$\cE^{(7+)}/Z^6$}\\[2pt]
 2 & -8.2x\,(4.1)   & -0.43x\,(22)    & -0.59x\,(30)    & 0.09x3\,(46)    & 0.37x\,(18)     & 0.35x\,(17)     & 0.31x\,(15)    \\
 3 & -9.5x\,(3.2)   & -0.72x\,(24)    & -0.89x\,(30)    & 0.06x3\,(21)    & 0.37x\,(12)     & 0.35x\,(12)     & 0.31x\,(10)    \\
 4 & -9.6x\,(2.4)   & -0.83x\,(21)    & -0.97x\,(24)    & 0.05x5\,(14)    & 0.31x2\,(78)    & 0.29x4\,(73)    & 0.26x1\,(65)   \\
 5 & -9.4x\,(1.9)   & -0.86x\,(17)    & -0.98x\,(20)    & 0.05x2\,(10)    & 0.26x6\,(53)    & 0.24x9\,(50)    & 0.21x9\,(44)   \\
 6 & -9.0x\,(1.5)   & -0.87x\,(14)    & -0.96x\,(16)    & 0.05x10\,(85)   & 0.23x0\,(38)    & 0.21x4\,(36)    & 0.18x7\,(31)   \\
 7 & -8.7x\,(1.2)   & -0.86x\,(12)    & -0.94x\,(13)    & 0.05x00\,(72)   & 0.20x2\,(29)    & 0.18x8\,(27)    & 0.16x2\,(23)   \\
 8 & -8.3x\,(1.0)   & -0.84x\,(11)    & -0.91x\,(11)    & 0.04x90\,(62)   & 0.18x1\,(23)    & 0.16x7\,(21)    & 0.14x4\,(18)   \\
 9 & -8.0x0\,(89)   & -0.82x5\,(91)   & -0.88x6\,(98)   & 0.04x77\,(54)   & 0.16x3\,(18)    & 0.15x0\,(17)    & 0.12x9\,(14)   \\
10 & -7.7x0\,(77)   & -0.80x6\,(80)   & -0.85x9\,(85)   & 0.04x63\,(48)   & 0.14x8\,(15)    & 0.13x6\,(14)    & 0.11x6\,(12)   \\
11 & -7.4x3\,(67)   & -0.78x6\,(71)   & -0.83x4\,(75)   & 0.04x50\,(43)   & 0.13x5\,(13)    & 0.12x4\,(12)    & 0.10x66\,(97)  \\
12 & -7.1x8\,(60)   & -0.76x8\,(63)   & -0.81x1\,(67)   & 0.04x35\,(39)   & 0.12x4\,(11)    & 0.11x32\,(99)   & 0.09x84\,(82)  
  \end{tabular}
\end{ruledtabular}
\end{table*}

%
%
%
\begin{table*}[htb]
\caption{Total theoretical ionization energies of $n=1$ and $n=2$ states in
  light helium-like ions, in cm$^{-1}$. $A$ is the nuclear mass number and
  $R_{\rm ch}$ is the root-mean-square nuclear charge radius. 
  \label{tab:total}}
\begin{ruledtabular}
  \begin{tabular}{ll....}
    $Z$  &  $A$&  \multicolumn{1}{c}{$R_{\rm ch}\ ${\rm [fm]}}& \multicolumn{1}{c}{$1^1S$}
    &  \multicolumn{1}{c}{$2^1S$}
    &  \multicolumn{1}{c}{$2^3S$}
    \\
    \hline\\[-5pt]
    2& 4& 1.6x76\,(3)   &     -198\,310.x665\,1\,(12)&     -32\,033.x228\,734\,(63)&     -38\,454.x694\,593\,(86)\\
    3& 7& 2.4x3\,(3)    &     -610\,078.x549\,(11)   &    -118\,704.x799\,71\,(79) &    -134\,044.x256\,96\,(98) \\
    4& 9& 2.5x2\,(1)    &  -1\,241\,256.x625\,(45)   &    -260\,064.x342\,7\,(38)  &    -284\,740.x785\,8\,(45)  \\
    5&11& 2.4x1\,(3)    &  -2\,091\,995.x58\,(13)    &    -456\,261.x994\,(12)     &    -490\,434.x928\,(14)   \\
    6&12& 2.4x70\,(2)   &  -3\,162\,423.x60\,(32)    &    -707\,370.x691\,(31)     &    -751\,130.x806\,(34)   \\
    7&14& 2.5x58\,(7)   &  -4\,452\,723.x93\,(66)    & -1\,013\,458.x475\,(66)     & -1\,066\,874.x012\,(72)   \\
    8&16& 2.7x01\,(6)   &  -5\,963\,074.x2\,(1.2)    & -1\,374\,588.x68\,(13)      & -1\,437\,720.x48\,(14)    \\
    9&19& 2.8x98\,(2)   &  -7\,693\,708.x5\,(2.1)    & -1\,790\,837.x48\,(22)      & -1\,863\,745.x75\,(24)    \\
    10&20& 3.0x05\,(2)   & -9\,644\,843.x7\,(3.5)    & -2\,262\,278.x68\,(36)      & -2\,345\,025.x61\,(39)    \\
    11&23& 2.9x94\,(2)   &-11\,816\,821.x8\,(5.4)    & -2\,789\,016.x99\,(57)      & -2\,881\,669.x22\,(60)    \\
    12&24& 3.0x57\,(2)   &-14\,209\,915.x2\,(8.1)    & -3\,371\,143.x82\,(86)      & -3\,473\,772.x54\,(90)    
  \end{tabular}
  \begin{tabular}{l....}
    $Z$  &   \multicolumn{1}{c}{$2^1P_1$}
    &  \multicolumn{1}{c}{$2^3P_0$} 
    &  \multicolumn{1}{c}{$2^3P_1$} 
    &  \multicolumn{1}{c}{$2^3P_2$}
    \\
    \hline\\[-5pt]
    2&        -27\,175.x771\,929\,(13)&     -29\,222.x838\,110\,(54)&     -29\,223.x826\,028\,(51)&     -29\,223.x902\,466\,(45)\\
    3&       -108\,270.x881\,302\,(69)&    -115\,812.x954\,89\,(40) &    -115\,818.x148\,68\,(38) &    -115\,816.x058\,08\,(34)\\
    4&       -243\,787.x567\,65\,(25) &    -257\,876.x174\,4\,(14)  &    -257\,887.x732\,4\,(14)  &    -257\,872.x840\,8\,(12)\\
    5&       -434\,000.x421\,29\,(74) &    -455\,041.x300\,3\,(37)  &    -455\,057.x499\,0\,(35)  &    -455\,004.x840\,0\,(31)\\
    6&       -679\,022.x761\,7\,(18)  &    -707\,232.x019\,2\,(81)  &    -707\,244.x528\,3\,(76)  &    -707\,108.x731\,2\,(66)\\
    7&       -978\,929.x118\,5\,(39)  & -1\,014\,453.x504\,(15)     & -1\,014\,444.x829\,(14)     & -1\,014\,153.x830\,(12)\\
    8&    -1\,333\,765.x988\,7\,(74)  & -1\,376\,741.x630\,(27)     & -1\,376\,682.x831\,(25)     & -1\,376\,131.x273\,(22)\\
    9&    -1\,743\,582.x370\,(13)     & -1\,794\,154.x597\,(44)     & -1\,794\,003.x382\,(41)     & -1\,793\,045.x585\,(35)\\
    10&   -2\,208\,407.x741\,(22)     & -2\,266\,761.x636\,(70)     & -2\,266\,460.x989\,(65)     & -2\,264\,903.x347\,(54)\\
    11&   -2\,728\,303.x815\,(35)     & -2\,794\,653.x26\,(11)      & -2\,794\,130.x721\,(97)     & -2\,791\,723.x350\,(82)\\
    12&   -3\,303\,295.x620\,(52)     & -3\,377\,923.x51\,(15)      & -3\,377\,091.x31\,(14)      & -3\,373\,518.x82\,(12)
  \end{tabular}
\end{ruledtabular}
\end{table*}

%
%

\begin{table*}[htb]
\caption{Comparison of theoretical and experimental transition energies.
Units are MHz for He and Li$^+$ and cm$^{-1}$ for other ions. 
Results by Drake are from 2005 for He \cite{drake:05:springer}, 
from 1994 for Li$^+$ \cite{riis:94}, and from 1988 for other ions
\cite{drake:88:cjp}. 
\label{tab:transen}} 
\begin{ruledtabular}
  \begin{tabular}{rlllc}
    $Z$ 
    & \multicolumn{1}{c}{This work}
    & \multicolumn{1}{c}{Drake}
    & \multicolumn{1}{c}{Experiment}
    & \multicolumn{1}{c}{Reference}           \\
    \hline\\[-5pt]
    \multicolumn{5}{l}{
      $2^3P_0$--$2^3S_1$ transition:
    }\\
    2 & 276\,764\,094.7(3.0) & 276\,764\,099(17)
    & 276\,764\,094.678\,8(21) & \cite{pastor:04}\\ 
    3 & 546\,560\,686(32)& 546\,560\,627 
    & 546\,560\,683.07(42)& \cite{riis:94}\\
    4 & 26\,864.6114(47)&
    26\,864.64(3) & 26\,864.612\,0(4)   & \cite{scholl:93}\\
    5 &  35\,393.628(14) &
    35\,393.70(8) & 35\,393.627(13)   & \cite{dinneen:91}\\
    8 &  60\,978.85(14)&
    60\,979.6(5)  & 60\,978.44(52)    & \cite{peacock:84}\\
    10&  78\,263.98(39)&
    78\,265.9(1.2)& 78\,265.0(1.2)    & \cite{peacock:84}\\
    \\[-5pt]
    \multicolumn{5}{l}{
      $2^3P_1$--$2^3S_1$ transition:
    }\\
    2 &  276\,734\,477.7(3.0) & 276\,734\,476(17)
    & 276\,764\,477.724\,2(20)  & \cite{pastor:04}\\ 
    3 &  546\,404\,980(31) & 546\,404\,885
    & 546\,404\,978.80(51) & \cite{riis:94}\\
    4 &  26\,853.0534(47)&
    26\,852.04(3) & 26\,853.053\,4(3)     & \cite{scholl:93}\\
    5 &  35\,377.429(14) &
    35\,377.40(8) & 35\,377.424(13)     & \cite{dinneen:91}\\
    8 &  61\,037.65(14) &
    61\,037.7(5)  & 61\,037.62(93)      & \cite{peacock:84}\\
    \\[-5pt]
    \multicolumn{5}{l}{
      $2^3P_2$--$2^3S_1$ transition:
    }\\
    2 &   276\,732\,186.1(2.9) & 276\,732\,183(17)
    & 276\,732\,186.593\,(15) & \cite{pastor:04}\\ 
    3 &   546\,467\,655(31) & 546\,467\,553
    & 546\,467\,657.21(44)  & \cite{riis:94}\\
    4 &  26\,867.9450(47)&
    26\,867.92(3) & 26\,867.948\,4(3)     & \cite{scholl:93}\\
    5 &  35\,430.088(14) &
    35\,430.02(8) & 35\,430.084(9)      & \cite{dinneen:91}\\
    8 &  61\,589.21(14)&
    61\,589.0(5)  & 61\,589.70(53)      & \cite{peacock:84}\\
    10&  80\,122.3(4)  &
    80\,121.6(1.2)& 80\,121.53(64)      & \cite{hallet:93}\\
    \\[-5pt]
    \multicolumn{5}{l}{
      $2^1P_1$--$2^1S_0$ transition:
    }\\
    4 &  16\,276.775(4)&
    16\,276.77(3) & 16\,276.774(9) & \cite{scholl:89}\\
    \\[-5pt]
    \multicolumn{5}{l}{
      $2^3P_1$--$2^1S_0$ transition:
    }\\
    7 &  986.36(7) & 986.6(3) & 986.3180(7) & \cite{thompson:98} \\
  \end{tabular}
\end{ruledtabular}
\end{table*}


\end{document}